\documentclass[aps,pra,twocolumn,nobibnotes,superscriptaddress,nofootinbib,10pt]{revtex4-1} 
\usepackage[latin1]{inputenc}
\usepackage{amsmath,amssymb}
\usepackage{mathrsfs} 
\usepackage{graphicx,color,colortbl}
\usepackage{rotating,array,tabularx,booktabs}
\usepackage{varwidth,xcolor}
\usepackage{placeins}
\usepackage{siunitx}
\usepackage[normalem]{ulem}%

\DeclareGraphicsExtensions{.pdf,.PDF,.png,.PNG}
\allowdisplaybreaks

\newcommand{\dif}{\mathrm{d}}%
\newcommand{\abs}[1]{\lvert#1\rvert}%
\newcommand{\norm}[1]{\lVert#1\rVert}%
\newcommand{\ZT}[1]{\textquotedblleft#1\textquotedblright}%
\newcommand{\ww}{\vec{u}}%
\newcommand{\ws}{u}%

\begin{document}

\title{Acoustic propulsion of nano- and microcones: dependence on particle size, acoustic energy density, and sound frequency}

\author{Johannes Vo\ss{}}
\affiliation{Institut f\"ur Theoretische Physik, Center for Soft Nanoscience, Westf\"alische Wilhelms-Universit\"at M\"unster, 48149 M\"unster, Germany}

\author{Raphael Wittkowski}
\email[Corresponding author: ]{raphael.wittkowski@uni-muenster.de}
\affiliation{Institut f\"ur Theoretische Physik, Center for Soft Nanoscience, Westf\"alische Wilhelms-Universit\"at M\"unster, 48149 M\"unster, Germany}

\begin{abstract}
Employing acoustofluidic simulations, we study the propulsion of cone-shaped nano- and microparticles by a traveling ultrasound wave. In particular, we investigate how the acoustic propulsion of the particles depends on their size and the energy density and frequency of the ultrasound wave. Our results reveal that the flow field generated around the particles depends on all three of these parameters. The results also show that the propulsion velocity of a particle increases linearly with the particle size and energy density and that an increase of the sound frequency leads to an increase of the propulsion velocity for frequencies below about $1\,\mathrm{MHz}$ but to a decrease of the propulsion velocity for larger frequencies. These findings are compared with preliminary results from the literature. 
\begin{figure}[htb]
\centering
\fbox{\includegraphics[width=8cm]{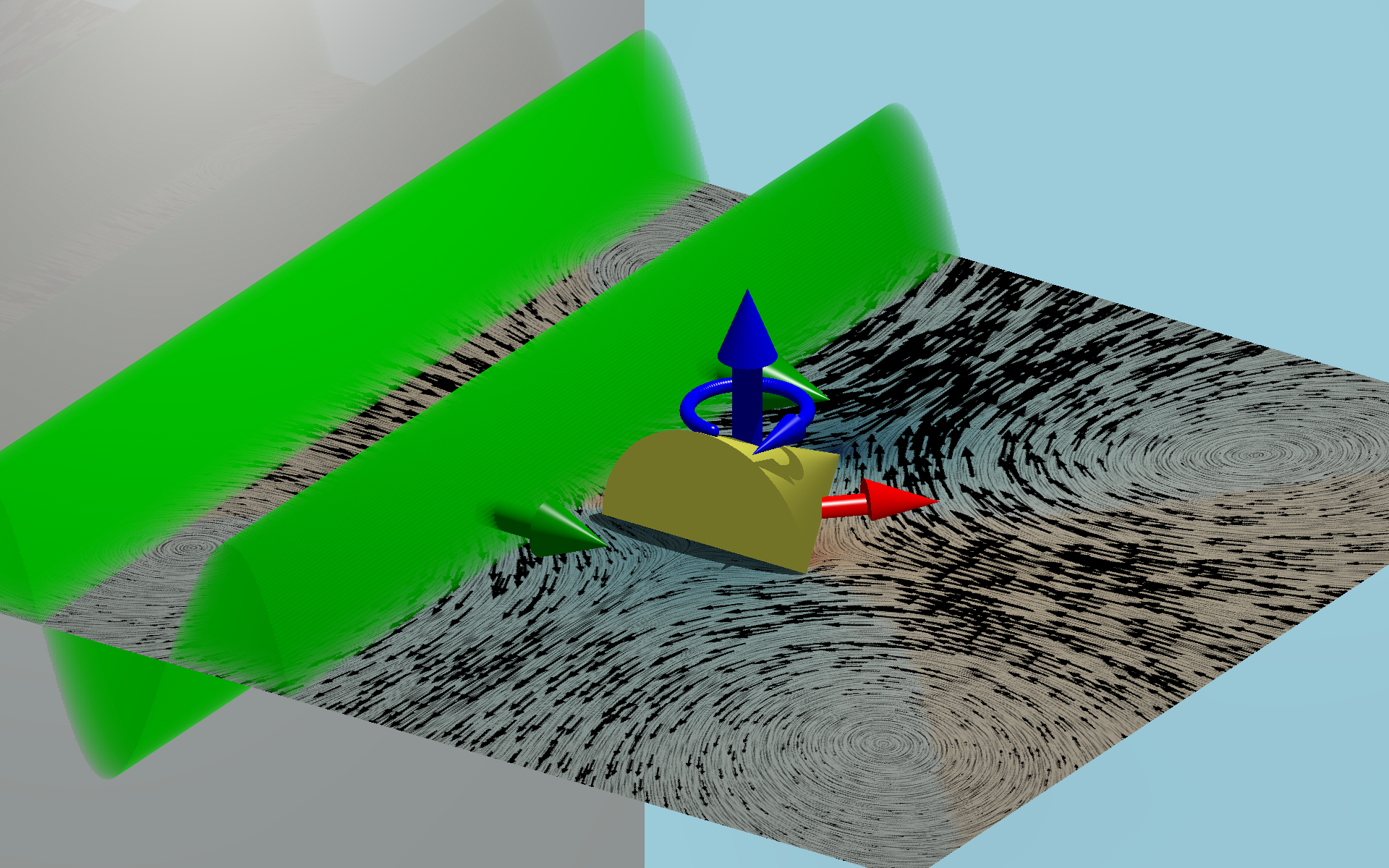}}%
\end{figure}
\vskip2ex\noindent\textbf{Keywords:} nano- and microcones, acoustic propulsion, size dependence, acoustic energy density dependence, frequency dependence, ultrasound
\end{abstract}
\maketitle

\section{Introduction}
Research on motile, artificial nano- and microparticles has resulted in a large number of different realizations of such particles \cite{BechingerdLLRVV2016,XuXZ2017,Venugopalan2020,FernandezRMHSS2020,YangEtAl2020}. 
They cover various propulsion mechanisms: chemical propulsion \cite{GaoDTLGZW2015,EstebanFernandezdeAvilaALGZW2018,OuEtAl2021}, light propulsion \cite{XuanSGWDH2018}, X-ray propulsion \cite{XuCLFPLK2019}, acoustic propulsion \cite{WangCHM2012,GarciaGradillaEtAl2013,AhmedEtAl2013,WuEtAl2014,WangLMAHM2014,GarciaGradillaSSKYWGW2014,BalkEtAl2014,AhmedGFM2014,NadalL2014,BalkEtAl2014,XuEtAl2015,FengYC2015,AhmedLNLSMCH2015,XieEtAl2015a,XieEtAl2015b,EstebanFernandezdeAvilaMSLRCVMGZW2015,Kiristi2015,WuEtAl2015a,WuEtAl2015b,RaoLMZCW2015,EstebanEtAl2016,KaynakONNLCH2016,SotoWGGGLKACW2016,KimGLZF2016,AhmedWBGHM2016,AhmedBJPDN2016,LaubliSAN2017,AhmedDHN2017,UygunEtAl2017,KaynakONLCH2017,CollisCS2017,EstebanFernandezEtAl2017,ZhouYWDW2017,ZhouZWW2017,ChenEtAl2018,HansenEtAl2018,SabrinaTABdlCMB2018,WangGWSGXH2018,EstebanEtAl2018,Zhou2018,LuSZWPL2019,QualliotineEtAl2019,GaoLWWXH2019,BeltranEtAl2019,BhuyanDBSGB2019,RenEtAl2019,ValdezLOESSWG2020,VossW2020,AghakhaniYWS2020,LiuR2020,DumyJMBGMHA2020,AghakhaniYWS2020,McneillNBM2020,McneillSWOLNM2021,MohantyEtAl2021,LiMMOP2021,VossW2021,VossW2022orientation}, and others \cite{EstebanFernandezdeAvilaALGZW2018,SafdarKJ2018,PengTW2017,KaganBCCEEW2012,XuanSGWDH2018,XuCLFPLK2019,FernandezRMHSS2020}. Among these mechanisms, acoustic propulsion has some significant advantages, such as that it is fuel-free and biocompatible and that it allows for supplying the particles continuously with energy \cite{XuGXZW2017,WangLWXLGM2019}.
With these properties, acoustically propelled nano- and microparticles have potential future applications in medicine \cite{LiEFdAGZW2017,PengTW2017,SotoC2018,WangGZLH2020,WangZ2021,Leal2021}, where they could be used, e.g., for drug delivery \cite{LuoFWG2018,ErkocYCYAS2019,WangZ2021,NitschkeW2021}, in materials science \cite{Visser2007,KummelSLVB2015,VanderMeerDF2016,JeanneretPKP2016,NeedlemanD2017,WangDZLGYLWM2019,RamananarivoDP2019,FratzlFKS2021}, where they could form active materials with exceptional properties \cite{JunH2010,McdermottKDKWSV2012}, and in other fields \cite{OphausKGT2021,teVrugtJW2021,WangJWZ2021,ShivalkarEtAl2021}. 
Although this type of particles has been intensively investigated in recent years, mainly based on experiments \cite{WangCHM2012,GarciaGradillaEtAl2013,AhmedEtAl2013,WuEtAl2014,WangLMAHM2014,GarciaGradillaSSKYWGW2014,BalkEtAl2014,AhmedGFM2014,AhmedLNLSMCH2015,LILXKLWW2015,WangDZSSM2015,EstebanFernandezdeAvilaMSLRCVMGZW2015,WuEtAl2015a,WuEtAl2015b,EstebanEtAl2016,SotoWGGGLKACW2016,AhmedWBGHM2016,KaynakONNLCH2016,UygunEtAl2017,KaynakONLCH2017,EstebanFernandezEtAl2017,RenZMXHM2017,ZhouYWDW2017,HansenEtAl2018,RenWM2018,SabrinaTABdlCMB2018,AhmedBJPDN2016,ZhouZWW2017,WangGWSGXH2018,EstebanEtAl2018,Zhou2018,TangEtAl2019,QualliotineEtAl2019,GaoLWWXH2019,RenEtAl2019,AghakhaniYWS2020,LiuR2020,ValdezLOESSWG2020,DumyJMBGMHA2020} but also using computer simulations \cite{KaynakONNLCH2016,AhmedBJPDN2016,SabrinaTABdlCMB2018,Zhou2018,WangGWSGXH2018,TangEtAl2019,RenEtAl2019,VossW2020,VossW2021,VossW2022orientation} and analytical approaches \cite{NadalL2014,CollisCS2017}, we are still at the beginning to understand the properties of these particles. 
For example, it is still rather unclear how their propulsion depends on the size of the particles, the energy density of the ultrasound, and its frequency.  
 
A few studies have addressed the dependence of the acoustic propulsion on the particle size so far \cite{NadalL2014,SotoWGGGLKACW2016,CollisCS2017}, but there is a discrepancy between experimental and theoretical work. 
In one experimental study \cite{SotoWGGGLKACW2016}, half-sphere cups (nanoshells) with different diameters were investigated in a standing ultrasound wave and a decreasing propulsion speed was observed for an increasing particle diameter. 
According to theoretical approaches \cite{NadalL2014,CollisCS2017}, however, the speed should increase with the particle size. 
In one study, a linear increase with the particle diameter was found for a sphere-like particle \cite{NadalL2014}. 
Another study found for different particle shapes a nonlinear but still increasing dependence of the propulsion on the particle size \cite{CollisCS2017}. 
However, with regard to future applications of acoustically propelled particles in medicine, the particle size is a critical parameter whose effect on the propulsion should be understood, since, depending on the drug-release method, the particles must have a sufficiently large volume \cite{SubjakovaOH2021,JinYDWWZ2021} or surface \cite{HansenEtAl2018,KimGLF2018,SubjakovaOH2021} but are not allowed to be so large that they can clog veins \cite{LuoFWG2018}.
 
The dependence of the acoustic propulsion on the energy density of the ultrasound has been studied in some experiments so far \cite{GarciaGradillaEtAl2013,WuEtAl2015a,AhmedWBGHM2016,KaynakONNLCH2016,AhmedBJPDN2016,KaynakONLCH2017,ZhouZWW2017,ZhouYWDW2017,UygunJSUW2016,Zhou2018,WangGWSGXH2018,BhuyanDBSGB2019,WangGZLH2020,McneillSWOLNM2021}, but there is not yet any investigation of this dependence that is based on simulations or analytical approaches. 
This constitutes a problem since in experimental work only the driving voltage applied to the ultrasound transducer is tuned directly, whereas the corresponding acoustic energy density that is established in the experimental setup near to the particles is not measured. So far, only a rough estimate $E\propto V^2$ \cite{Bruus2012} can be used to convert the driving voltage $V$ into the energy density $E$. In reality, the dependence of these quantities is affected by many details of the experimental setup, such as the acoustic coupling of adjacent components between the transducer and the particles. 
The existing experimental studies found either a linear \cite{WuEtAl2015a,KaynakONLCH2017,UygunJSUW2016,ZhouYWDW2017,Zhou2018,BhuyanDBSGB2019} or a quadratic \cite{GarciaGradillaEtAl2013,AhmedWBGHM2016,KaynakONNLCH2016,AhmedBJPDN2016,ZhouZWW2017,WangGWSGXH2018,WangGZLH2020} dependence of the propulsion speed on the driving voltage. Therefore, it now needs to be clarified which of these scaling relations is the correct one and what is the actual dependence of the propulsion speed on the acoustic energy density.
Determining these relations will be helpful to foresee the speed of acoustically propelled particles in particular applications, such as in medicine where the ultrasound intensity has to be limited to ensure harmlessness \cite{BarnettEtAl2000}. 

The frequency of the ultrasound is hard to change in most experiments since they use standing instead of traveling ultrasound waves \cite{WangCHM2012,AhmedEtAl2013,GarciaGradillaEtAl2013,GarciaGradillaSSKYWGW2014,WangLMAHM2014,AhmedGFM2014,BalkEtAl2014,WuEtAl2014,WuEtAl2015a,EstebanFernandezdeAvilaMSLRCVMGZW2015,XuEtAl2015,Kiristi2015,RaoLMZCW2015,SotoWGGGLKACW2016,AhmedWBGHM2016,EstebanEtAl2016,KaynakONNLCH2016,ZhouYWDW2017,KaynakONLCH2017,ZhouZWW2017,EstebanFernandezEtAl2017,UygunEtAl2017,XuXZ2017,Zhou2018,SabrinaTABdlCMB2018,HansenEtAl2018,WangGWSGXH2018,EstebanEtAl2018,GaoLWWXH2019,QualliotineEtAl2019,TangEtAl2019,LuSZWPL2019,BhuyanDBSGB2019,DumyJMBGMHA2020,LiuR2020,ValdezLOESSWG2020,WangGZLH2020,McneillSWOLNM2021}. 
Therefore, little is known about the dependence of the propulsion on the frequency. 
Up to now, there are only one experimental work \cite{AhmedBJPDN2016} and two analytical studies \cite{NadalL2014,CollisCS2017} that addressed this dependence. 
The experimental study \cite{AhmedBJPDN2016} observed a local minimum of the propulsion and even a switch of the propulsion direction when changing the frequency. 
In contrast, the analytical studies \cite{NadalL2014,CollisCS2017} found that the propulsion velocity increases linearly with the frequency. 
A better understanding of the dependence of the propulsion on the frequency would also be helpful for predicting the propulsion speed of a particle in a particular application. For example, in medical applications, the ultrasound frequency cannot be chosen arbitrarily, since the penetration depth of the ultrasound in biological tissue depends strongly on the frequency \cite{Wells1975}.
 
In this work, we, therefore, study carefully how the acoustic propulsion of nano- and microparticles depends on the particle size, acoustic energy density, and sound frequency. 
We consider cone-shaped particles, which are known to have a relatively efficient acoustic propulsion and are thus particularly relevant for applications \cite{VossW2020,VossW2021,VossW2022orientation}, and an ultrasound field consisting of a planar traveling ultrasound wave, which is more application relevant than the frequently chosen standing ultrasound waves \cite{VossW2020,VossW2021,VossW2022orientation}.
To calculate the sound-induced flow field that is generated around a particle and the resulting propulsion force and torque that act on the particle, we use acoustofluidic simulations.

\section{\label{results}Results and discussion}
We study the time-averaged flow field generated around a cone-shaped particle in water, which has diameter $\sigma$ and height $h=\sigma$ and is exposed to a planar traveling ultrasound wave, and the corresponding time-averaged propulsion force and torque that are exerted on the particle in the stationary state, for various values of the particle's size $\sigma$, the ultrasound's pressure amplitude $\Delta p$ (and thus energy density $E$), and the ultrasound's frequency $f$.   
The water is initially at standard temperature and standard pressure and quiescent.
Each parameter is varied separately from the other parameters, which are then fixed to their reference values $\sigma_\mathrm{R}=2^{-1/2}\SI{}{\micro\metre}$, $\Delta p_\mathrm{R}=\SI{10}{\kilo\pascal}$ (i.e., $E_\mathrm{R}=\SI{22.7}{\milli\joule\,\metre^{-3}}$), and $f_\mathrm{R}=\SI{1}{\MHz}$. 
See Methods for details.

\subsection{Dependence on particle size}
First, we study the dependence of the particle's flow field and propulsion force and torque on the particle's size $\sigma$.
We vary the particle's diameter as $\sigma\in[0.1,10]\sigma_\mathrm{R}$ while keeping the pressure amplitude at $\Delta p = \Delta p_\mathrm{R}$ and the frequency at $f=f_\mathrm{R}$.

\subsubsection{\label{sec:flowfieldsize}Flow field}
\begin{figure*}[htb]
\centering
\includegraphics[width=\linewidth]{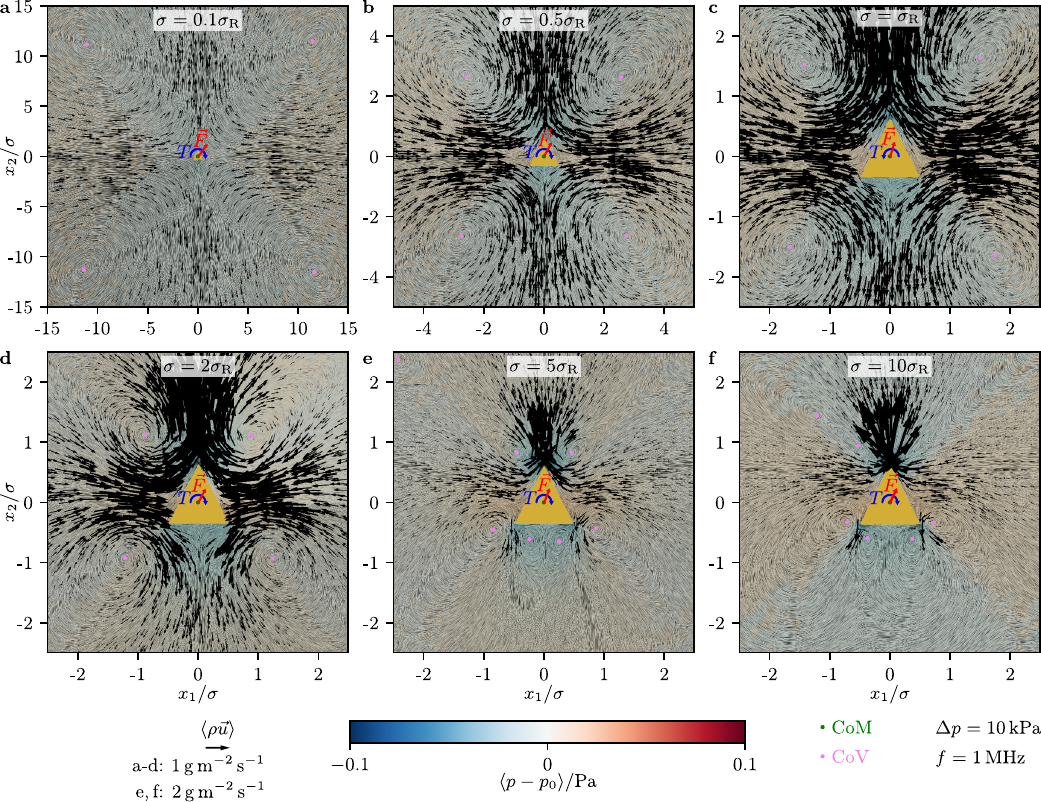}%
\caption{\label{fig:fig1}The time-averaged mass-current density $\langle\rho\ww\rangle$ and reduced pressure $\langle p-p_{0}\rangle$ for all considered particle diameters $\sigma$. A violet dot represents the center of a vortex (CoV) in the flow field and a green dot represents the center of mass (CoM) of the particle.
The directions of the propulsion force $\vec{F}$ and propulsion torque $T$ are indicated.}
\end{figure*}

The simulation results for the flow field are shown in Fig.\ \ref{fig:fig1}.
When the particle's diameter increases, we see that the structure of the flow field around the particle changes significantly.
For small diameters, the flow field is qualitatively similar to the structure that has been reported in Ref.\ \cite{VossW2020,VossW2021,VossW2022orientation}. 
When $\sigma=0.1\sigma_\mathrm{R}$, the flow field is dominated by four large vortices at the top left, top right, bottom left, and bottom right of the particle.
Their centers have similar distances from the center of mass of the particle and form the edges of a square. 
As a consequence of the assembly of vortices, the fluid flows with similar strength on the left and right towards the particle and below and above the particle away from it.
Moreover, the pressure is increased on the left and right and decreased below and above the particle. 

When $\sigma$ is increased up to $\sigma=2\sigma_\mathrm{R}$, the structure of the flow field remains qualitatively similar, but the centers of the vortices slowly diverge from the center of mass of the particle. 
Thereby, the upper two vortex centers become closer to each other than the lower ones so that the vortex assembly changes from a square to an isosceles trapezoid. 
The overall strength of the fluid flow increases. 
While the strength of the flow remains similar on the left and right of the particle, it becomes weaker than the lateral flow below the particle and stronger than the lateral flow above the particle. 
Hence, the strongest fluid flow occurs in front of the particle between the two upper vortices. 
With increasing $\sigma$, there thus occurs an increasing asymmetry of the fluid flow in front of and behind the particle. 
Furthermore, the minima and maxima of the pressure field become more pronounced when $\sigma$ increases. 
Figure \ref{fig:fig1}\textbf{c} shows the fluid flow for the reference case with $\sigma=\sigma_\mathrm{R}$, $\Delta p = \Delta p_\mathrm{R}$, and $f=f_\mathrm{R}$.

These trends continue when $\sigma$ increases beyond $\sigma=2\sigma_\mathrm{R}$. 
However, the structure of the fluid flow changes significantly for larger $\sigma$.
When proceeding to $\sigma=5\sigma_\mathrm{R}$, two secondary vortices occur between the two (primary) lower vortices. 
Furthermore, far away from the particle at its top left, a further secondary vortex center occurs. 
Thus, the mirror symmetry of the assembly of vortices is broken. 

When increasing $\sigma$ further to $\sigma=10\sigma_\mathrm{R}$, the centers of the secondary vortices approach the centers of the neighboring primary vortices. 
Moreover, the primary vortex at the top right of the particle disappears, which makes the vortex assembly even more asymmetric.  

\begin{figure*}[htb]
\centering
\includegraphics[width=\linewidth]{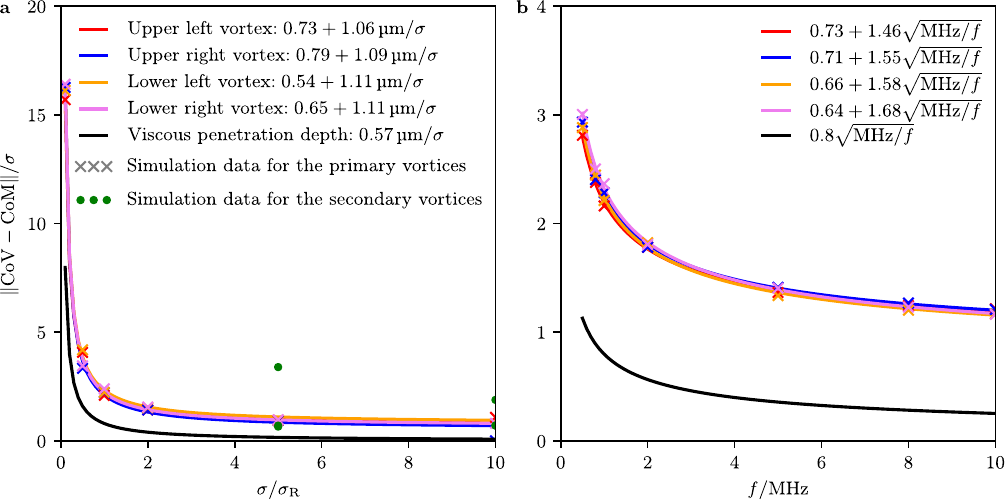}%
\caption{\label{fig:fig2}Simulation data (symbols) and fit functions (colored curves) for the distances $\norm{\mathrm{CoV}-\mathrm{CoM}}$ of the centers of the vortices in the flow field from the center of mass of the particle for varying \textbf{a} particle diameter $\sigma$ and \textbf{b} ultrasound frequency $f$. 
For comparison, the viscous penetration depth (black curve) is also shown.}
\end{figure*}

We now study the dependence on the particle size of the distance of the vortex centers from the center of mass of the particle in more detail. 
Figure \ref{fig:fig2}\textbf{a} shows these distances as a function of the particle's diameter $\sigma$. 
One can see that, in terms of $\sigma$, the distances rapidly decline when $\sigma$ increases. 
The dependence of the distances $\delta_\mathrm{pv}$ of the primary vortices on $\sigma$ can be described by a function 
\begin{align}
\delta_\mathrm{pv}(\sigma)=a\sigma + b\,\SI{}{\micro\metre}
\label{eq:deltapv}%
\end{align}
with coefficients $a$ and $b$. 
The latter coefficient becomes dominant for small particle sizes and determines a minimal vortex-to-particle distance that is not deceeded for any value of $\sigma$. 

When function \eqref{eq:deltapv} is fitted to the simulation data for the vortex-to-particle distances, one obtains fit functions that are in excellent agreement with the simulation data. 
The resulting fit functions are stated as equations and visualized in Fig.\ \ref{fig:fig2}\textbf{a}. 
For the second term in Eq.\ \eqref{eq:deltapv}, we find $b\,\SI{}{\micro\metre} \approx \SI{1.09}{\micro\metre} \approx 2 \delta_\mathrm{vpd}$, which is approximately the thickness of the viscous boundary layer in which vortices form near a boundary (Schlichting streaming) that is exposed to ultrasound \cite{Boluriaan2003}.
Here, $\delta_\mathrm{vpd}$ is the viscous penetration depth \cite{LandauL1987,WiklundRO2012} 
\begin{equation}
\delta_\mathrm{vpd}=\sqrt{\frac{\nu_\mathrm{s}}{\pi\rho_0 f}}\approx\SI{0.57}{\micro\metre} ,
\label{eq:deltavpd}%
\end{equation}
where for our situation $\nu_\mathrm{s}=\SI{1.002}{\milli\pascal\,\second}$ is the shear viscosity and $\rho_0=\SI{998}{\kilogram\,\metre^{-3}}$ is the mean mass density of the fluid and $f=f_\mathrm{R}$ is the frequency of the ultrasound.
For comparison, $\delta_\mathrm{vpd}$ is also shown in Fig.\ \ref{fig:fig2}\textbf{a}.

\subsubsection{\label{sec:propulsionsize}Propulsion force and torque}
\begin{figure*}[tb]
\centering
\includegraphics[width=\linewidth]{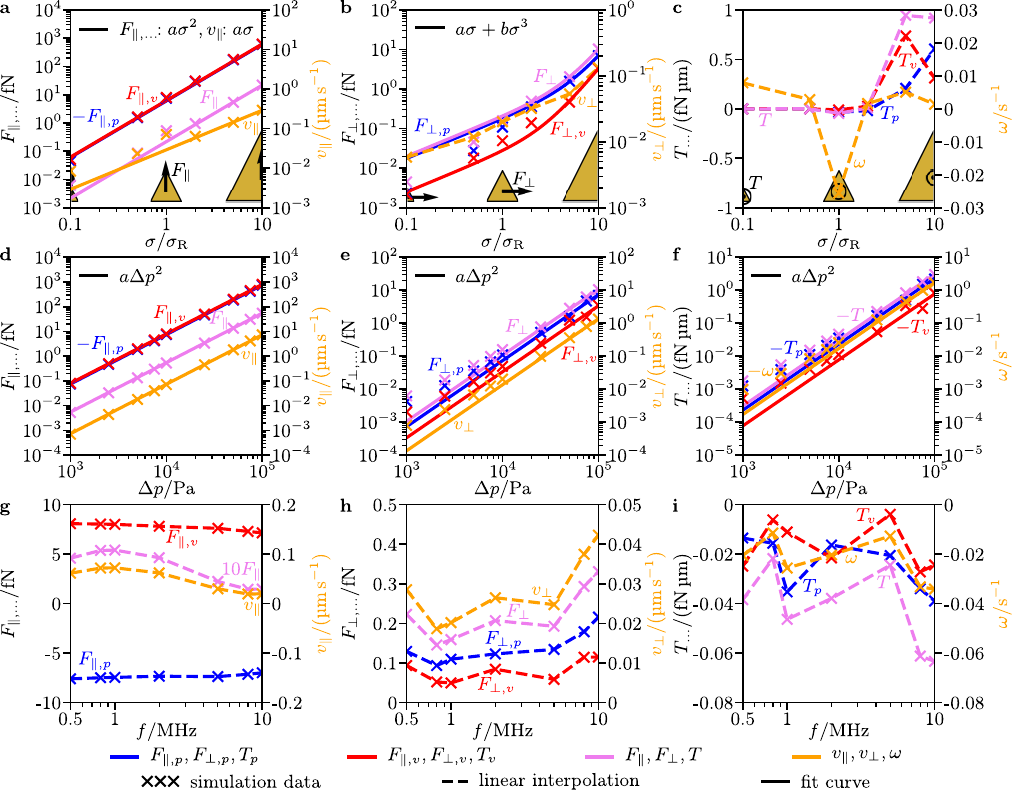}%
\caption{\label{fig:fig3}Simulation data and linear interpolation or fit curve for the time-averaged forces $F_{\parallel}$ and $F_{\perp}$ and torque $T$ acting on the particle, their pressure components $F_{\parallel,p}$, $F_{\perp, p}$, and $T_p$ and viscous components $F_{\parallel, v}$, $F_{\perp, v}$, and $T_v$, and the corresponding translational velocities $v_\parallel$ and $v_\perp$ and angular velocity $\omega$ of the particle for various \textbf{a}-\textbf{c} particle diameters $\sigma$, \textbf{d}-\textbf{f} ultrasound pressure amplitudes $\Delta p$, and \textbf{g}-\textbf{i} ultrasound frequencies $f$. 
Some curves show the negative of a quantity to avoid negative values in a logarithmic plot or a multiple of a quantity for better visibility.}
\end{figure*}

The simulation results for the propulsion force and torque are shown in Fig.\ \ref{fig:fig3}\textbf{a}-\textbf{c} (see Methods for the definitions of the components of the propulsion force and torque).
%

One can see that the propulsion force $F_\parallel$ parallel to the particle's orientation increases with the particle's diameter $\sigma$ from $F_\parallel=\SI{6.30}{\cdot10^{-3}\,\femto\newton}$ to $F_\parallel=\SI{21.81}{\femto\newton}$. 
Its pressure component $F_{\parallel,p}$ decreases from $F_{\parallel,p}=\SI{-4.49}{\cdot10^{-2}\,\femto\newton}$ to $F_{\parallel,p}=\SI{-595.32}{\femto\newton}$, whereas its viscous component $F_{\parallel,v}$ increases from $F_{\parallel,v}=\SI{5.12}{\cdot10^{-2}\,\femto\newton}$ to $F_{\parallel,v}=\SI{617.13}{\femto\newton}$.
The translational propulsion velocity $v_\parallel$ parallel to the particle's orientation increases from $v_\parallel=\SI{8.40}{\cdot10^{-3}\,\micro\metre\,\second^{-1}}$ to $v_\parallel=\SI{2.92}{\cdot10^{-1}\,\micro\metre\,\second^{-1}}$.  
This increase of the propulsion speed is in line with the observation of Section \ref{sec:flowfieldsize} that the flow near the particle becomes stronger when $\sigma$ increases.

Next, we consider the perpendicular components of the propulsion.
We see that they all increase with the particle's diameter $\sigma$.
The propulsion force $F_\perp$ perpendicular to the particle's orientation increases from $F_\perp=\SI{4.52}{\cdot10^{-3}\,\femto\newton}$ to $F_\perp=\SI{10.26}{\femto\newton}$, its pressure component $F_{\perp,p}$ increases from $F_{\perp,p}=\SI{2.32}{\cdot10^{-3}\,\femto\newton}$ to $F_{\perp,p}=\SI{6.97}{\femto\newton}$, its viscous component $F_{\perp,v}$ increases from $F_{\perp,v}=\SI{2.20}{\cdot10^{-3}\,\femto\newton}$ to $F_{\perp,v}=\SI{3.29}{\femto\newton}$, and the translational propulsion velocity $v_\perp$ perpendicular to the particle's orientation increases from $v_\perp=\SI{5.85}{\cdot10^{-3}\,\micro\metre\,\second^{-1}}$ to $v_\perp=\SI{1.33}{\cdot10^{-1}\,\micro\metre\,\second^{-1}}$.

When considering the angular components of the propulsion, the dependence on the particle size is more complicated.
The propulsion torque $T$ increases from $T=\SI{7.04}{\cdot10^{-6}\,\femto\newton\,\micro\metre}$ to $T=\SI{9.45}{\cdot10^{-1}\,\femto\newton\,\micro\metre}$ at $\sigma=5\sigma_\mathrm{R}$ and slightly decreases to $T=\SI{9.17}{\cdot10^{-1}\,\femto\newton\,\micro\metre}$ afterward. 
Its pressure component $T_p$ increases from $T_p=\SI{9.59}{\cdot10^{-6}\,\femto\newton\,\micro\metre}$ to $T_p=\SI{6.05}{\cdot10^{-1}\,\femto\newton\,\micro\metre}$, whereas the viscous component $T_v$ increases from $T_v=\SI{-2.55}{\cdot10^{-6}\,\femto\newton\,\micro\metre}$ to $T_v=\SI{7.37}{\cdot 10^{-1}\,\femto\newton\,\micro\metre}$ at $\sigma=5\sigma_\mathrm{R}$ and then decreases to $T_v=\SI{3.11}{\cdot10^{-1}\,\femto\newton\,\micro\metre}$.
For the angular propulsion velocity $\omega$ we find first a decrease from $\omega=\SI{7.89}{\cdot10^{-3}\,\second^{-1}}$ to $\omega=\SI{-2.56}{\cdot10^{-2}\,\second^{-1}}$ at $\sigma=\sigma_\mathrm{R}$, then an increase to $\omega=\SI{5.07}{\cdot10^{-3}\,\second^{-1}}$ at $\sigma=5\sigma_\mathrm{R}$, and finally a decrease to $\omega=\SI{1.40}{\cdot10^{-3}\,\second^{-1}}$. 
The values of $\omega$ are small for all considered particle sizes and can be neglected compared to rotational Brownian motion. 
In the case $\sigma=\sigma_\mathrm{R}$, where $\abs{\omega}$ is maximal, its value corresponds to a rotation by 90\textdegree{} within approximately $\SI{60}{\second}$. 
In contrast, the time scale for a reorientation of the particle by Brownian motion is $D_{\mathrm{R}}^{-1}=\SI{0.43}{\second}$ with the particle's rotational diffusion coefficient $D_{\mathrm{R}}=\SI{2.34}{\second^{-1}}$. 

For some of the curves, we can state simple fit functions: 
\begin{align}
\mathfrak{f}(\sigma)=
\begin{cases}
a \sigma^2, & \text{for }\mathfrak{f}\in\{F_{\parallel,p}, F_{\parallel,v}, F_\parallel\}, \\
a \sigma, & \text{for }\mathfrak{f}\in\{v_{\parallel}\}, \\
a \sigma + b \sigma^3, & \text{for }\mathfrak{f}\in\{F_{\perp,p}, F_{\perp,v}, F_\perp\}.
\end{cases}\label{eq:fit_size} 
\end{align}
The values of the fit coefficients $a$ and $b$ are given in Tab.\ \ref{tab:tab2}.
\begin{table*}[tb]
\centering
\caption{\label{tab:tab2}Quantities for which a fit curve is shown in Fig.\ \ref{fig:fig3}, their exact relations to the other quantities, their fit functions \eqref{eq:fit_size} and \eqref{eq:fit_energy_density}, and the values of the corresponding fit coefficients $a$ and $b$.
The dependence of the translational resistance coefficient $\mathrm{K}_{22}(\sigma)$ on the particle diameter $\sigma$ follows from Eq.\ \eqref{eq:Ksigma}.}
\begin{ruledtabular}
\begin{tabular}{lcccc}
\textbf{Quantity} &\textbf{Relation} &\textbf{Fit function}  &$\boldsymbol{a}$  & $\boldsymbol{b}$\\
\hline
$F_{\parallel,p}(\sigma)$ & --- & $a\sigma^2$  & $\SI{-12.020}{\femto\newton\,\micro\metre^{-2}}$ & --- \\
$F_{\parallel,v}(\sigma)$ & --- & $a\sigma^2$ & $\SI{12.456}{\femto\newton\,\micro\metre^{-2}}$ & --- \\
$F_{\parallel}(\sigma)$ & --- & $a\sigma^2$  & $\SI{4.361}{\cdot10^{-1}\,\femto\newton\,\micro\metre^{-2}}$ & --- \\
$v_{\parallel}(\sigma)$ & $F_{\parallel}(\sigma)/(\nu_\mathrm{s}\mathrm{K_{22}}(\sigma))$ & $a\sigma$ & $\SI{4.122}{\cdot10^{-2}\,\second^{-1}}$  & --- \\
$F_{\perp,p}(\sigma)$ & --- & $a\sigma+b\sigma^3$  &$\SI{2.490}{\cdot10^{-1}\,\femto\newton\micro\metre^{-1}}$ & $\SI{1.476}{\cdot10^{-2}\,\femto\newton\,\micro\metre^{-3}}$ \\
$F_{\perp,v}(\sigma)$ & --- & $a\sigma+b\sigma^3$  & $\SI{3.539}{\cdot10^{-2}\,\femto\newton\,\micro\metre^{-1}}$ & $\SI{8.584}{\cdot10^{-3}\,\femto\newton\,\micro\metre^{-3}}$ \\
$F_{\perp}(\sigma)$ & --- & $a\sigma+b\sigma^3$  & $\SI{2.844}{\cdot10^{-1}\,\femto\newton\,\micro\metre^{-1}}$ & $\SI{2.334}{\cdot10^{-2}\,\femto\newton\,\micro\metre^{-3}}$ \\
\hline
$F_{\parallel,p}(\Delta p)$ & --- & $a\Delta p^2$  & $\SI{-7.467}{\cdot10^{-2}\,\femto\newton\,\kilo\pascal^{-2}}$ & --- \\
$F_{\parallel,v}(\Delta p)$ & --- & $a\Delta p^2$  & $\SI{8.015}{\cdot10^{-2}\,\femto\newton\,\kilo\pascal^{-2}}$ & --- \\
$F_{\parallel}(\Delta p)$ & ---& $a\Delta p^2$  & $\SI{5.480}{\cdot10^{-3}\femto\newton\,\kilo\pascal^{-2}}$ & --- \\
$v_{\parallel}(\Delta p)$ & $F_{\parallel}(\Delta p)/(\nu_\mathrm{s}\mathrm{K_{22}})$ & $a\Delta p^2$  & $\SI{7.325}{\cdot10^{-4}\,\micro\metre\,\second^{-1}\kilo\pascal^{-2}}$ & --- \\
$F_{\perp,p}(\Delta p)$ & --- & $a\Delta p^2$ & $\SI{7.316}{\cdot10^{-4}\,\femto\newton\,\kilo\pascal^{-2}}$ & --- \\
$F_{\perp,v}(\Delta p)$ & --- & $a\Delta p^2$  & $\SI{3.366}{\cdot10^{-4}\,\femto\newton\,\kilo\pascal^{-2}}$ & --- \\
$F_{\perp}(\Delta p)$ & ---& $a\Delta p^2$  & $\SI{1.068}{\cdot10^{-3}\,\femto\newton\,\kilo\pascal^{-2}}$ & --- \\
$v_{\perp}(\Delta p)$ & $(-\mathrm{\Omega_{33}}F_{\perp}(\Delta p)+\mathrm{C_{31}}T(\Delta p))/(\nu_\mathrm{s}(\mathrm{C^2_{31}}-\mathrm{K_{11}}\mathrm{\Omega_{33}}))$ & $a\Delta p^2$  & $\SI{1.358}{\cdot10^{-4}\,\micro\metre\,\second^{-1}\,\kilo\pascal^{-2}}$ & --- \\
$T_{p}(\Delta p)$ & --- & $a\Delta p^2$  & $\SI{-2.299}{\cdot10^{-4}\,\femto\newton\,\micro\metre\,\kilo\pascal^{-2}}$ & --- \\
$T_{v}(\Delta p)$ & --- & $a\Delta p^2$  & $\SI{-7.698}{\cdot10^{-5}\,\femto\newton\,\micro\metre\,\kilo\pascal^{-2}}$& --- \\
$T(\Delta p)$ & ---& $a\Delta p^2$  & $\SI{-3.069}{\cdot10^{-4}\,\femto\newton\,\micro\metre\,\kilo\pascal^{-2}}$ & --- \\
$\omega(\Delta p)$ & $(\mathrm{C_{31}}F_{\perp}(\Delta p)-\mathrm{K_{11}}T(\Delta p))/(\nu_\mathrm{s}(\mathrm{C^2_{31}}-\mathrm{K_{11}}\mathrm{\Omega_{33}}))$ & $a\Delta p^2$  & $\SI{-1.696}{\cdot10^{-4}\,\second^{-1}\,\kilo\pascal^{-2}}$ & ---  
\end{tabular} 
\end{ruledtabular}%
\end{table*}
These fit functions are in good agreement with the simulation data.
The linear fit function for $v_{\parallel}(\sigma)$ results from the fit function for $F_{\parallel}(\sigma)$ and the Stokes law \eqref{eq:velocity} (see Tab.\ \ref{tab:tab2}).  
   
We can compare our findings with the results of previous studies. 
Reference \cite{NadalL2014}, which is based on an analytical approach, found $v_\parallel(\sigma)\propto \sigma$, which implies
$F_\parallel(\sigma)\propto K_{22}(\sigma) v_\parallel(\sigma) \propto \sigma^2$ with $K_{22}(\sigma)\propto \sigma$.
These scaling relations are in line with our fit functions, although the system considered in Ref.\ \cite{NadalL2014} differs significantly from the system that is considered in our article.
In particular, Ref.\ \cite{NadalL2014} considered near-sphere particles and a standing ultrasound wave.  
  
In Ref.\ \cite{CollisCS2017}, $v_\parallel$ is studied for dumbbell-shaped particles in a standing ultrasound wave by an analytical approach. 
According to this reference, $v_\parallel$ should scale as $v_\parallel\propto-\beta^{5/2}$ for $\beta \ll 1$ and as $v_\parallel\propto\beta^{1/2}$ for $\beta\gg 1$.
Furthermore, there should be a sign change of $v_\parallel$ at $\beta\in O(1)$. 
Here, $\beta$ is the acoustic Reynolds number $\beta=\pi\rho_0\sigma^2 f/(2\nu_\mathrm{s})$ and ranges between $0.2$ and $78$ when we vary $\sigma$.
The scaling for $\beta \gg 1$ is in very good agreement with our simulation results and the different scaling for smaller values of $\beta$ is in line with the fact that our simulation results differ somewhat from the linear scaling for $\sigma<2\sigma_\mathrm{R}$.
Since $v_\parallel\propto\sigma$ for $\beta\gg 1$, the propulsion speed is constantly at 0.04 body lengths per second for a large particle size. 

According to Ref.\ \cite{Bruus2012acoustofluidics}, the force $F_{\perp}$ has two contributions. 
One is the acoustic radiation force, which scales as $\propto\sigma^3$, and the other one is the acoustic streaming force, which scales as $\propto\sigma$. 
This provides an explanation for why our fit function for $F_{\perp}(\sigma)$ requires two terms.
However, both components act in opposite directions according to Ref.\ \cite{Bruus2012acoustofluidics} so there should be a switch in the propulsion direction for smaller values of $\sigma$, whereas the prefactors of both terms in our fit function have the same sign. 
This discrepancy might result from numerical inaccuracies of the simulations that affect mainly the results for small particles. 

Reference \cite{SotoWGGGLKACW2016}, which is based on experiments, found a decrease of $v_\parallel$ for increasing particle size. 
This is in contrast to our findings, but the discrepancy is likely to originate from the differences of the performed experiments compared to the system studied in the present work.
In Ref.\ \cite{SotoWGGGLKACW2016}, half-sphere-cup-shaped particles (nanoshells) in a standing ultrasound wave were studied and it is known that at least the particle shape has a significant effect on the propulsion \cite{VossW2020}. 
Furthermore, the frequency of the ultrasound ($f=\SI{2.66}{\MHz}$) was different from the frequency that we considered when we varied the particle size.

\subsection{Dependence on acoustic energy density}
Second, we study the dependence of the particle's flow field and propulsion force and torque on the ultrasound's acoustic energy density $E$. For this purpose, we vary the ultrasound's pressure amplitude as $\Delta p \in[0.1,10]\Delta p_\mathrm{R}$ and thus the acoustic energy density as $E \in[0.01,100] E_\mathrm{R}$ while keeping the particle diameter at $\sigma=\sigma_\mathrm{R}$ and the frequency at $f=f_\mathrm{R}$.

\subsubsection{\label{sec:flowfieldenergydensity}Flow field}
\begin{figure*}[htb]
\centering
\includegraphics[width=\linewidth]{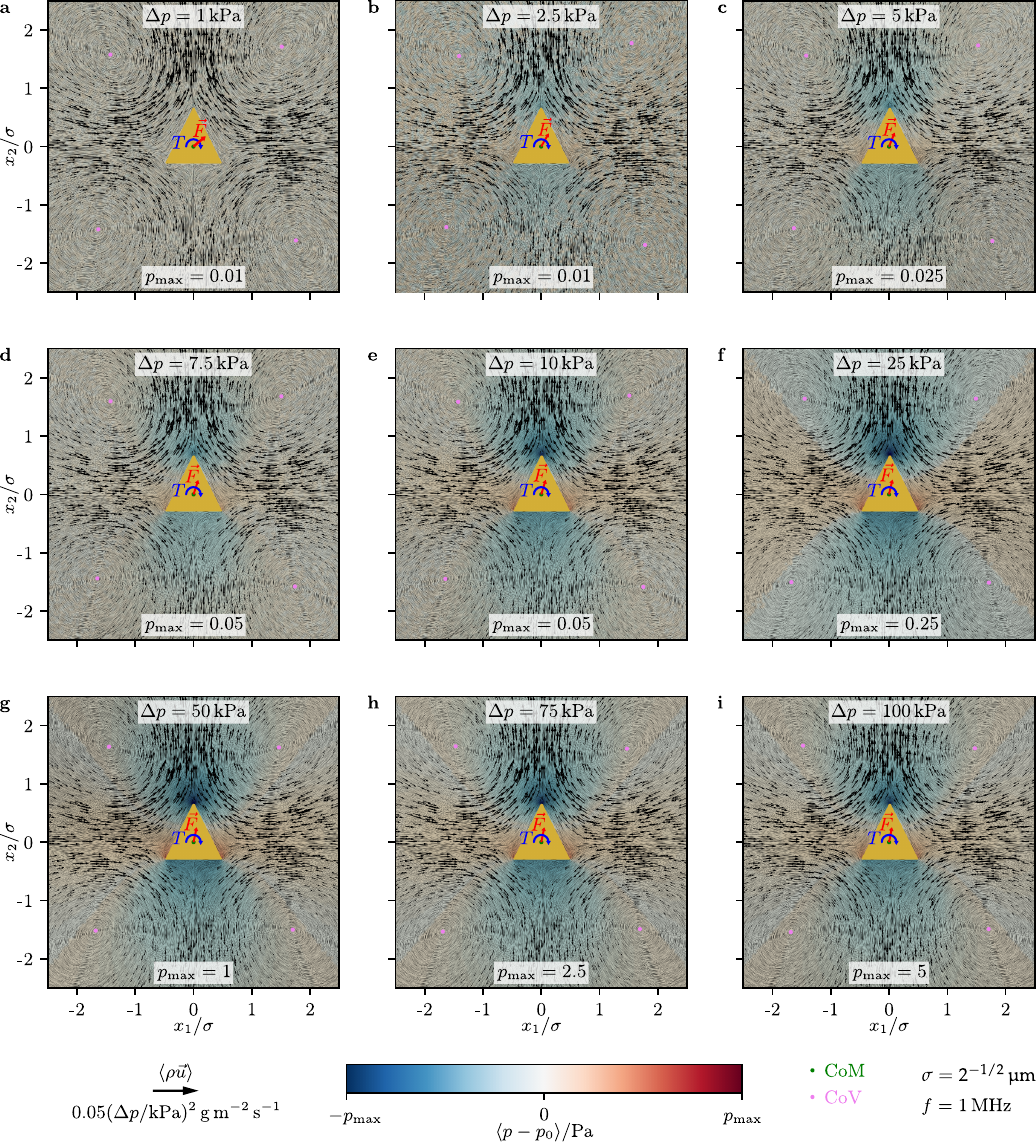}%
\caption{\label{fig:fig4}The same as Fig.\ \ref{fig:fig1}, but now for the considered pressure amplitudes $\Delta p$ and thus energy densities of the ultrasound. In each plot, the value of $p_\mathrm{max}$ denotes the chosen maximum pressure on the color bar.}
\end{figure*}

The simulation results for the flow field are shown in Fig.\ \ref{fig:fig4}.
When the ultrasound's pressure amplitude increases, we see that the flow field remains qualitatively similar to the reference case that is shown in Fig.\ \ref{fig:fig1}\textbf{c}. 
Again, the flow field is dominated by four vortices at the top left, top right, bottom left, and bottom right of the particle.
The positions of the centers of the vortices are approximately independent of the pressure amplitude $\Delta p$. 
Their distances from the center of mass of the particle are $\approx\SI{2.16}{}\sigma$ for the top left, $\approx\SI{2.21}{}\sigma$ for the top right, $\approx\SI{2.25}{}\sigma$ for the bottom left, and $\approx\SI{2.34}{}$ for the bottom right vortex. 
However, the fluid flow becomes stronger and the minimum and maximum of the pressure field become more pronounced when $\Delta p$ increases.
The strength of the mass-current density approximately scales as $\propto \Delta p^2$ and for $\Delta p=\SI{100}{\kilo\pascal}$, the minimum pressure occurs at the upper tip of the particle whereas the maximum pressure occurs near the lower end of the left and right edges of the particle.

\subsubsection{\label{sec:propulsionenergydensity}Propulsion force and torque}
The simulation results for the propulsion force and torque are shown in Fig.\ \ref{fig:fig3}\textbf{d}-\textbf{f}.
One can see that, except for their signs, the propulsion force $F_{\parallel}$ parallel to the particle's orientation, its pressure component $F_{\parallel,p}$ and viscous component $F_{\parallel, v}$, 
the translational propulsion velocity $v_\parallel$ parallel to the particle's orientation, 
the propulsion force $F_{\perp}$ perpendicular to the particle's orientation, its pressure component $F_{\perp, p}$ and viscous component $F_{\perp, v}$, 
the translational propulsion velocity $v_\perp$ perpendicular to the particle's orientation, 
the propulsion torque $T$, its pressure component $T_p$ and viscous component $T_v$, 
and the angular propulsion velocity $\omega$ all scale similarly with the ultrasound's pressure amplitude $\Delta p$.
$F_{\parallel}$ increases from $F_\parallel=\SI{5.25}{\cdot10^{-3}\,\femto\newton}$ to $F_\parallel=\SI{54.76}{\femto\newton}$,
$F_{\parallel,p}$ decreases from $F_{\parallel,p}=\SI{-7.48}{\cdot10^{-2}\,\femto\newton}$ to $F_{\parallel,p}=\SI{-746.82}{}$,
$F_{\parallel, v}$ increases from $F_{\parallel,v}=\SI{8.00}{\cdot10^{-2}\,\femto\newton}$ to $F_{\parallel,v}=\SI{801.58}{\femto\newton}$,
$v_\parallel$ increases from $v_\parallel=\SI{7.01}{\cdot10^{-4}\,\micro\metre\,\second^{-1}}$ to $v_\parallel=\SI{73.21}{\micro\metre\,\second^{-1}}$,
$F_{\perp}$ increases from $F_\perp=\SI{6.25}{\cdot10^{-3}\,\femto\newton}$ to $F_\perp=\SI{10.49}{\femto\newton}$,
$F_{\perp, p}$ increases from $F_{\perp,p}=\SI{4.22}{\cdot10^{-3}\,\femto\newton}$ to $F_{\perp,p}=\SI{7.02}{\femto\newton}$,
$F_{\perp, v}$ increases from $F_{\perp,v}=\SI{2.03}{\cdot10^{-3}\,\femto\newton}$ to $F_{\perp,v}=\SI{3.47}{\femto\newton}$,
$v_\perp$ increases from $v_\perp=\SI{7.95}{\cdot10^{-4}\,\micro\metre\,\second^{-1}}$ to $v_\perp=\SI{1.33}{\micro\metre\,\second^{-1}}$,
$T$ decreases from $T=\SI{-1.79}{\cdot10^{-3}\,\femto\newton\,\micro\metre}$ to $T=\SI{-3.01}{\femto\newton\,\micro\metre}$,
$T_p$ decreases from $T_p=\SI{-1.26}{\cdot10^{-3}\,\femto\newton\,\micro\metre}$ to $T_p=\SI{-2.19}{\femto\newton\,\micro\metre}$,
$T_v$ decreases from $T_v=\SI{-5.25}{\cdot10^{-4}\,\femto\newton\,\micro\metre}$ to $T_v=\SI{-8.19}{\cdot10^{-1}\,\femto\newton\,\micro\metre}$,
and $\omega$ decreases from $\omega=\SI{-9.88}{\cdot10^{-4}\,\second^{-1}}$ to $\omega=\SI{-1.66}{\second^{-1}}$.
The increase of the propulsion speed $v_\parallel$ is in line with the observation of Section \ref{sec:flowfieldenergydensity} that the flow near the particle becomes stronger when $\Delta p$ increases.
Similar to our findings for a variation of the particle size (see Section \ref{sec:propulsionsize}), the torques are again small compared to Brownian motion. 
The maximum $\omega=\SI{-1.66}{\second^{-1}}$ of the angular velocity corresponds to a rotation of the particle by 90\textdegree{} within $\SI{1}{\second}$. 
In contrast, rotational Brownian motion reorients the particle already on a time scale of $D_{\mathrm{R}}^{-1}=\SI{0.43}{\second}$, where $D_{\mathrm{R}}=\SI{2.34}{\second^{-1}}$.   

For the dependence of all these quantities on $\Delta p$, a simple fit function can be given:
\begin{align}
&\mathfrak{g}(\Delta p) = a \Delta p ^2 = 2 a \rho_0 c_\mathrm{f}^2 E,\label{eq:fit_energy_density}\\ 
&\text{for }\mathfrak{g}\in\{F_{\parallel,p}, F_{\parallel,v}, F_\parallel, v_\parallel, 
F_{\perp,p}, F_{\perp,v}, F_{\perp}, v_{\perp},
T_{p}, T_{v}, T, \omega\}. \nonumber
\end{align}
Here, $c_\mathrm{f}$ is the speed of sound and the values of the fit coefficient $a$ are given in Tab.\ \ref{tab:tab2}.
The agreement of the fit functions for $F_\parallel$, $F_{\parallel,p}$, $F_{\parallel,v}$, and $v_\parallel$ with the simulation results is excellent. 
For the other quantities, the overall agreement is good and the agreement for large values of $\Delta p$ is even very good. 
%
The fit function for $v_{\parallel}(\Delta p)$ results from the fit function for $F_{\parallel}(\Delta p)$ and the Stokes law \eqref{eq:velocity}. 
Similarly, the fit functions for $v_{\perp}(\Delta p)$ and $\omega(\Delta p)$ result from the fit functions for $F_{\perp}(\Delta p)$ and $T(\Delta p)$, respectively. 
These analytical relations are explicitly given in Tab.\ \ref{tab:tab2}.
 
Again, we compare our findings with the results of previous studies. 
At first, we compare with theoretical studies. There, in line with our findings, a quadratic dependency of $v_\parallel$ on $\Delta p$ was found \cite{NadalL2014,CollisCS2017,SabrinaTABdlCMB2018}. 
Note that the largest pressure amplitude $\Delta p=\SI{100}{\kilo\pascal}$ that we considered in our simulations is so large that linearizations of the Navier-Stokes equations that are used in Refs.\ \cite{NadalL2014,CollisCS2017,SabrinaTABdlCMB2018} are not applicable.
According to Ref.\ \cite{Bruus2012acoustofluidics}, the acoustic radiation force and the acoustic streaming force that contribute to $F_{\perp}(\Delta p)$ both scale as $\propto \Delta p^2$. This is consistent with the scaling $F_{\perp}(\Delta p)\propto \Delta p^2$ of our corresponding fit function. 

A comparison with experimental results is not straightforwardly possible, since experimental studies only measure the amplitude of the voltage $V$ that drives the ultrasound transducer but not directly the pressure amplitude $\Delta p$ in the liquid near the particle. 
Therefore, we need to employ the rough estimate $E\propto V^2$ \cite{Bruus2012} to convert the driving voltage $V$ into the energy density $E\propto \Delta p^2$ and thus into $\Delta p$.
With this conversion, the scalings for $F_\parallel$ and $v_\parallel$ with $\Delta p$ that have been found in the experiment-based studies \cite{GarciaGradillaEtAl2013,AhmedWBGHM2016,KaynakONNLCH2016,AhmedBJPDN2016,ZhouZWW2017,WangGWSGXH2018,WangGZLH2020} are consistent with our results. 
Furthermore, Ref.\ \cite{ZhouZWW2017} suggests the same scaling of $T$ with $\Delta p$ based on experiments that we found in our simulations. 

In Fig.\ \ref{fig:fig3}\textbf{a}, we observed that the propulsion velocity $v_\parallel$ increases with the particle's diameter $\sigma$ and reaches a value of $v_\parallel=\SI{2.92}{\cdot10^{-1}\,\micro\metre\,\second^{-1}}$ for $\sigma=10\sigma_\mathrm{R}$.
Since we have revealed how $v_\parallel$ scales with $\Delta p$ and $E\propto\Delta p^2$, we can now determine how fast a particle with diameter $\sigma=10\sigma_\mathrm{R}$ would move if it is exposed to ultrasound with the maximum energy density $E_\mathrm{max}=\SI{4.9}{\joule\,\metre^{-3}}$ that is permitted by the U.S.\ Food and Drug Administration for diagnostic applications of ultrasound in the human body \cite{BarnettEtAl2000}. 
Rescaling $v_\parallel$ to this energy density results in a propulsion speed of approximately 9 body lengths per second, which is quite fast.

\begin{figure*}[htb]
\centering
\includegraphics[width=\linewidth]{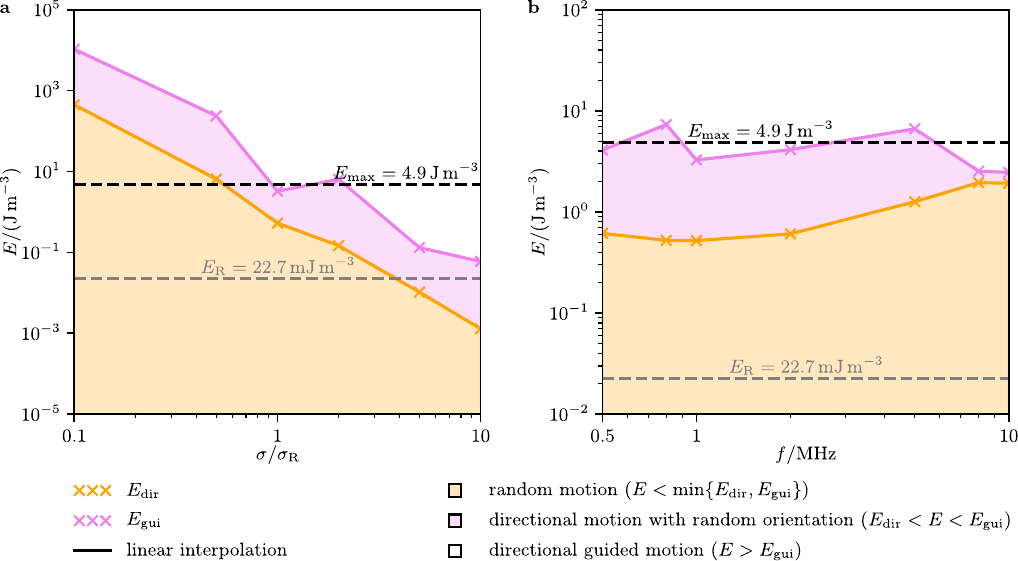}%
\caption{\label{fig:fig5}Dependence of the qualitative particle motion on the acoustic energy density $E$ and \textbf{(a)} particle diameter $\sigma$ or \textbf{(b)} ultrasound frequency $f$. 
The threshold energy densities $E_\mathrm{dir}$ for directional motion and $E_\mathrm{gui}$ for directional guided motion are calculated using Eqs.\ \eqref{eq:E_dir_lim} and \eqref{eq:E_Br_lim}, respectively. 
For comparison, also the reference acoustic energy density $E_\mathrm{R}$ (see Tab.\ \ref{tab:Parameters}) and the maximum permissible acoustic energy density $E_\mathrm{max}$ for diagnostic human ultrasound applications \cite{BarnettEtAl2000} are shown.}
\end{figure*}

Next, we assess how a free particle would move along a trajectory when it is observed for some time.  
The type of motion of the particle would depend on its translational propulsion velocity, angular propulsion velocity, and Brownian motion. 
Thus, it will depend on the particle's diameter $\sigma$ and the acoustic energy density $E$. 
Therefore, we now study how the particle's type of motion depends on $\sigma$ and $E$.
With the classification for the qualitative particle motion that is described in the Methods, 
the type of motion can be classified as \ZT{random motion} 
($E < \min\{E_\mathrm{dir},E_\mathrm{gui}\}$), 
where the particle's Brownian rotation dominates the particle's translational and rotational propulsion, 
\ZT{directional motion with random orientation} 
($E_\mathrm{dir} < E < E_\mathrm{gui}$), 
where the particle's translational propulsion dominates the particle's Brownian rotation and the Brownian rotation dominates the particle's rotational propulsion, 
or \ZT{directional guided motion} 
($E > E_\mathrm{gui}$), 
where the particle's translational propulsion and rotational propulsion dominate the Brownian rotation.
The energy density thresholds $E_\mathrm{dir}$ and $E_\mathrm{gui}$ are defined by Eqs.\ \eqref{eq:E_dir_lim} and \eqref{eq:E_Br_lim} in the Methods.  
These thresholds depend on the particle's diameter $\sigma$, the particle's rotational diffusion coefficient $D_\mathrm{R}$, which in turn depends on $\sigma$, and the translational and angular propulsion velocities $v_\parallel$ and $\omega$, which in turn depend on $\sigma$ and $E$. 
To determine the values of $v_\parallel$ and $\omega$ as functions of $\sigma$ and $E$, we use our simulation results for their dependence on $\sigma$ (see Section \ref{sec:propulsionsize}) and their proportionality to $E$ that we found in the present section.  
The results are shown in Fig.\ \ref{fig:fig5}\textbf{a}.
One can see that $E_\mathrm{gui}$ is always larger than $E_\mathrm{dir}$ and that, except for a small dip of $E_\mathrm{gui}$ at $\sigma=\sigma_\mathrm{R}$, both energy density thresholds decrease when $\sigma$ increases. A reason for this trend is that Brownian rotation decreases when the particle size increases. 
Thus, for larger particles, a lower intensity of the ultrasound is required to overcome random motion and to reach directional motion with random orientation or even directional guided motion.
For example, particles with $\sigma \gtrsim \sigma_\mathrm{R}$ can show directional motion for harmless ultrasound intensities ($E<E_\mathrm{max}$).
This means that for medical applications such as drug delivery, the particle size should not be significantly smaller than $\sigma_\mathrm{R}=2^{-1/2}\SI{}{\micro\metre}$.

\subsection{Dependence on sound frequency}
Third, we study the dependence of the particle's flow field and propulsion force and torque on the ultrasound's frequency $f$. We vary the frequency as $f \in[0.5,10]f_\mathrm{R}$ while keeping the particle diameter at $\sigma=\sigma_\mathrm{R}$ and the pressure amplitude at $\Delta p=\Delta p_\mathrm{R}$.

\subsubsection{\label{sec:flowfieldfrequency}Flow field}
\begin{figure*}[htb]
\centering
\includegraphics[width=\linewidth]{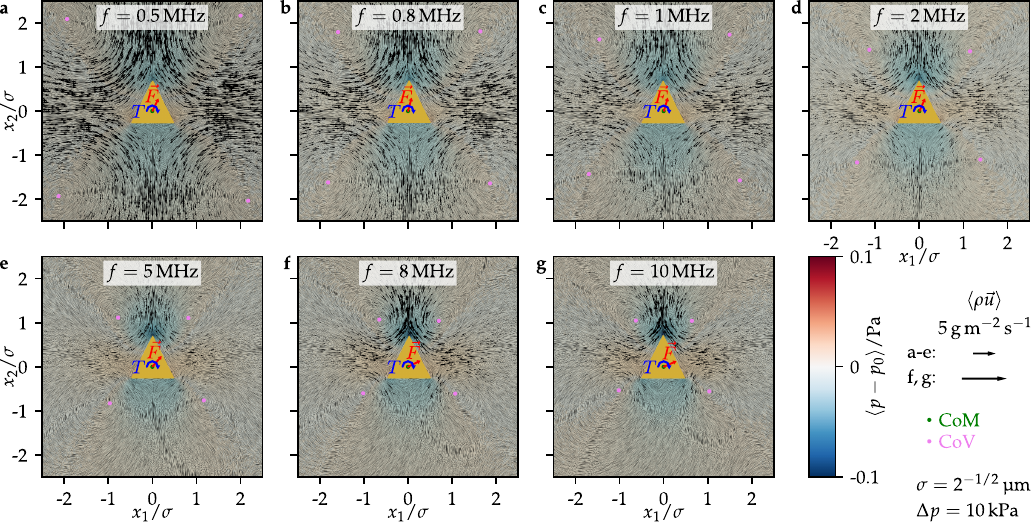}%
\caption{\label{fig:fig6}The same as Fig.\ \ref{fig:fig1}, but now for the considered frequencies $f$ of the ultrasound.}
\end{figure*}

The simulation results for the flow field are shown in Fig.\ \ref{fig:fig6}.
When the ultrasound's frequency $f$ increases, we see that the structure of the flow field around the particle changes significantly, but not as strongly as for an increase of the particle size (see Section \ref{sec:flowfieldsize}).
For all frequencies, the basic structure of the flow field is dominated by four vortices at the top left, top right, bottom left, and bottom right of the particle, similar to the reference case that is shown in Fig.\ \ref{fig:fig1}\textbf{c}. 

When $f$ increases, the centers of the vortices approach the center of mass of the particle. 
Thereby, the centers of the two upper vortices become closer than the centers of the two lower vortices. 
Thus, the assembly of the centers of the vortices again forms an isosceles trapezoid.
As a consequence of the change in the vortex assembly, the fluid flow does not retain a similar strength besides, below, and above the particle, but becomes much stronger in front of the particle than besides or below it.
The approach of the centers of the two upper vortices and the associated concentration of the fluid flow in front of the particle upon an increase of the frequency are similar to what we observed for a moderate increase of the particle size (see Section \ref{sec:flowfieldsize}).
 
However, a formation of new vortices and a disappearance of vortices, as we have seen in Section \ref{sec:flowfieldsize} for a strong increase of the particle size, do not occur in the present situation. 
Moreover, different from what we observed for an increase of $\sigma$ (see Section \ref{sec:flowfieldsize}) and $\Delta p$ (see Section \ref{sec:flowfieldenergydensity}), the strength of the fluid flow and the variation of the pressure field do not increase with $f$. 
In contrast, the mass-current density declines (except for a small area in front of the particle), and the pressure variation remains approximately constant when $f$ increases. 

We now consider the distances of the centers of the vortices from the center of mass of the particle in more detail. 
Figure \ref{fig:fig2}\textbf{b} shows how these distances depend on the ultrasound frequency $f$. 
One can see that the distances of the vortex centers from the particle's center of mass rapidly decrease when $f$ increases. 
The dependence of these distances $\delta_\mathrm{pv}$ on $f$ can be described by a function 
\begin{align}
\delta_\mathrm{pv}(f) = a\sigma + b\,\SI{}{\micro\metre} \sqrt{\SI{}{\MHz}/f} 
\label{eq:deltapvf}%
\end{align}
with coefficients $a$ and $b$. 
In Fig.\ \ref{fig:fig2}\textbf{b}, the functions that result from fitting Eq.\ \eqref{eq:deltapvf} to the simulation data for the vortex-to-particle distances are visualized. For these fit functions, also explicit equations are given. Their agreement with the simulation data is excellent. 
For the second term in Eq.\ \eqref{eq:deltapvf}, we obtain $b\,\SI{}{\micro\metre} \sqrt{\SI{}{\MHz}/f}\approx 1.57\sigma \sqrt{\SI{}{\MHz}/f}\approx 2\delta_\mathrm{vpd}$, which is the typical size of the boundary layer vortex thickness for the Schlichting streaming \cite{Boluriaan2003}. 
Remarkably, the scaling of $\delta_\mathrm{pv}$ with $f$ is in line with the $f$-dependence of the viscous penetration depth $\delta_\mathrm{vpd}(f) \approx 0.8\sigma\sqrt{\SI{}{\MHz}/f}$ (see Eq.\ \eqref{eq:deltavpd}).

\subsubsection{Propulsion force and torque}
The simulation results for the propulsion force and torque are shown in Fig.\ \ref{fig:fig3}\textbf{g}-\textbf{i}.
Interestingly, the dependence of $F_{\parallel}$, $F_{\parallel, p}$, $F_{\parallel, v}$, $v_\parallel$, $F_{\perp}$, $F_{\perp, p}$, $F_{\perp, v}$, $v_\perp$, $T$, $T_p$, $T_v$, and $\omega$ on the ultrasound's frequency $f$ is rather different and in the most cases quite complicated. 
$F_{\parallel}$, $F_{\parallel, p}$, $F_{\parallel, v}$, and $v_\parallel$ have the simplest frequency dependence of these.  
$F_{\parallel}$ first increases from $F_{\parallel}=\SI{4.58}{\cdot10^{-1}\,\femto\newton}$ to $F_{\parallel}=\SI{5.40}{\cdot10^{-1}\,\femto\newton}$ at $f=\SI{1}{\MHz}$ and then decreases to $F_{\parallel}=\SI{1.46}{\cdot10^{-1}\,\femto\newton}$; 
$F_{\parallel, p}$ increases from $F_{\parallel, p}=\SI{-7.61}{\femto\newton}$ to $F_{\parallel, p}=\SI{-6.99}{\femto\newton}$; 
$F_{\parallel, v}$ decreases from $F_{\parallel, v}=\SI{8.07}{\femto\newton}$ to $F_{\parallel, v}=\SI{7.14}{\femto\newton}$; 
$v_\parallel$ first increases from $v_\parallel=\SI{6.11}{\cdot10^{-2}\,\micro\metre\,\second^{-1}}$ to $v_\parallel=\SI{7.21}{\cdot10^{-2}\,\micro\metre\,\second^{-1}}$ at $f=1\,\mathrm{MHz}$ and then decreases to $v_\parallel=\SI{1.95}{\cdot10^{-2}\,\micro\metre\,\second^{-1}}$.
The frequency dependence of $F_{\perp}$, $F_{\perp, p}$, $F_{\perp, v}$, and $v_\perp$ is more complicated. 
$F_{\perp}$ first decreases from $F_{\perp}=\SI{2.23}{\cdot10^{-1}\,\femto\newton}$ to $F_{\perp}=\SI{1.45}{\cdot10^{-1}\,\femto\newton}$ at $f=\SI{0.8}{\MHz}$, then increases to $F_{\perp}=\SI{2.07}{\cdot10^{-1}\,\femto\newton}$ at $f=\SI{2}{\MHz}$, afterward decreases to $F_{\perp}=\SI{1.93}{\cdot10^{-1}\,\femto\newton}$ at $f=\SI{5}{\MHz}$, and finally increases to $F_{\perp}=\SI{3.31}{\cdot10^{-1}\,\femto\newton}$; 
$F_{\perp, p}$ first decreases from $F_{\perp, p}=\SI{1.29}{\cdot10^{-1}\,\femto\newton}$ to $F_{\perp, p}=\SI{9.33}{\cdot10^{-2}\,\femto\newton}$ at $f=\SI{0.8}{\MHz}$ and then increases to $F_{\perp, p}=\SI{2.16}{\cdot10^{-1}\,\femto\newton}$;  
$F_{\perp, v}$ first decreases from $F_{\perp, v}=\SI{9.36}{\cdot10^{-2}\,\femto\newton}$ to $F_{\perp, v}=\SI{4.94}{\cdot10^{-2}\,\femto\newton}$ at $f=\SI{1}{\MHz}$, then increases to $F_{\perp, v}=\SI{8.41}{\cdot10^{-2}\,\femto\newton}$ at $f=\SI{2}{\MHz}$, afterward decreases to $F_{\perp, v}=\SI{5.85}{\cdot10^{-2}\,\femto\newton}$ at $f=\SI{5}{\MHz}$, and finally increases to $F_{\perp, v}=\SI{1.15}{\cdot10^{-1}\,\femto\newton}$;  
$v_\perp$ first decreases from $v_\perp=\SI{2.84}{\cdot10^{-2}\,\micro\metre\,\second^{-1}}$ to $v_\perp=\SI{1.86}{\cdot10^{-2}\,\micro\metre\,\second^{-1}}$ at $f=\SI{0.8}{\MHz}$, then increases to $v_\perp=\SI{2.64}{\cdot10^{-2}\,\micro\metre\,\second^{-1}}$ at $f=\SI{2}{\MHz}$, afterward decreases to $v_\perp=\SI{2.46}{\cdot10^{-2}\,\micro\metre\,\second^{-1}}$ at $f=\SI{5}{\MHz}$, and finally increases to $v_\perp=\SI{4.22}{\cdot10^{-2}\,\micro\metre\,\second^{-1}}$.
Also, the frequency dependence of $T$, $T_p$, $T_v$, and $\omega$ is rather complicated, but now increases and decreases are typically inverted compared to the previous case.
$T$ first increases from $T=\SI{-3.84}{\cdot10^{-2}\,\femto\newton\,\micro\metre}$ to $T=\SI{-2.18}{\cdot10^{-2}\,\femto\newton\,\micro\metre}$ at $f=\SI{0.8}{\MHz}$, then decreases to $T=\SI{-4.64}{\cdot10^{-2}\,\femto\newton\,\micro\metre}$ at $f=\SI{1}{\MHz}$, afterward increases to $T=\SI{-2.45}{\cdot10^{-2}\,\femto\newton\,\micro\metre}$ at $f=\SI{5}{\MHz}$, and finally decreases to $T=\SI{-6.34}{\cdot10^{-2}\,\femto\newton\,\micro\metre}$;  
$T_p$ first decreases from $T_p=\SI{-1.35}{\cdot10^{-2}\,\femto\newton\,\micro\metre}$ to $T_p=\SI{-3.53}{\cdot10^{-2}\,\femto\newton\,\micro\metre}$ at $f=\SI{1}{\MHz}$, then increases to $T_p=\SI{-1.63}{\cdot10^{-2}\,\femto\newton\,\micro\metre}$ at $f=\SI{2}{\MHz}$, and finally decreases to $T_p=\SI{-3.90}{\cdot10^{-2}\,\femto\newton\,\micro\metre}$;  
$T_v$ first increases from $T_v=\SI{-2.49}{\cdot10^{-2}\,\femto\newton\,\micro\metre}$ to $T_v=\SI{-6.18}{\cdot10^{-3}\,\femto\newton\,\micro\metre}$ at $f=\SI{0.8}{\MHz}$, then decreases to $T_v=\SI{-2.16}{\cdot10^{-2}\,\femto\newton\,\micro\metre}$ at $f=\SI{2}{\MHz}$, afterward increases to $T_v=\SI{-3.92}{\cdot10^{-3}\,\femto\newton\,\micro\metre}$ at $f=\SI{5}{\MHz}$, then decreases again to $T_v=\SI{-2.72}{\cdot10^{-2}\,\femto\newton\,\micro\metre}$ at $f=\SI{8}{\MHz}$, and finally increases to $T=\SI{-2.44}{\cdot10^{-2}\,\femto\newton\,\micro\metre}$; 
and $\omega$ first increases from $\omega=\SI{-2.04}{\cdot10^{-2}\,\second^{-1}}$ to $\omega=\SI{-1.14}{\cdot10^{-2}\,\second^{-1}}$ at $f=\SI{0.8}{\MHz}$, then decreases to $\omega=\SI{-2.55}{\cdot10^{-2}\,\second^{-1}}$ at $f=\SI{1}{\MHz}$, afterward increases to $\omega=\SI{-1.26}{\cdot10^{-2}\,\second^{-1}}$ at $f=\SI{5}{\MHz}$, and finally decreases to $\omega=\SI{-3.39}{\cdot10^{-2}\,\second^{-1}}$. 
The overall decrease of the propulsion speed is in line with the observation of Section \ref{sec:flowfieldfrequency} that the overall flow in the liquid around the particle declines when $f$ increases.
Since the values of $\omega$ are of the same order of magnitude as for a variation of the particle size (see Section \ref{sec:propulsionsize}), the angular propulsion can again be neglected compared to rotational Brownian motion. 
%
Due to the complicated frequency dependence of these quantities, we cannot provide simple fit functions.

Next, we compare our results with the literature. 
The observed overall decrease of the propulsion speed $v_\parallel$ for increasing frequency $f$ is in contradiction to the results of an analytical approach that is described in Ref.\ \cite{CollisCS2017}.
According to this reference, the propulsion speed should increase with the frequency. 
However, the results of this reference are not directly applicable to our situation.
They require an acoustic Reynolds number $\beta$ with $\beta \ll 1$ or $\beta\gg 1$, whereas in our system $\beta\in [0.4,7.8]$.
This is likely to explain the discrepancy between their results and our findings. 
The observed overall increase of the force $F_{\perp}$ with $f$ is consistent with the increasing force that a traveling ultrasound wave exerts on a spherical particle in the direction of ultrasound propagation \cite{Doinikov1994,SettnesB2012}. 

Finally, we address the type of motion that a free particle would exhibit and its dependence on the frequency $f$ and acoustic energy density $E$. 
The corresponding results are shown in Fig.\ \ref{fig:fig5}\textbf{b}.
Different from what we observed for a variation of the particle diameter $\sigma$ (see Section \ref{sec:propulsionenergydensity}), the energy density $E_\mathrm{dir}$, which constitutes an upper threshold for random motion of the particle, increases with $f$ and the energy density $E_\mathrm{gui}>E_\mathrm{dir}$, which constitutes an upper threshold for directional motion with random orientation and a lower threshold for directional guided motion, fluctuates around the energy density $E_\mathrm{max}$ and only slightly decreases when $f$ increases.  
Hence, for all considered frequencies, directional motion of the particle is possible for harmless ultrasound intensities ($E<E_\mathrm{max}$).

\section{\label{conclusions}Conclusions}
In this work, we carefully investigated how the acoustic propulsion of cone-shaped nano- and microparticles by a planar traveling ultrasound wave depends on the size of the particles, the energy density of the ultrasound, and its frequency. 
We found that all three parameters have a strong influence on the flow field generated around the nano- and microcones and their resulting propulsion velocity. 
When increasing the particle size or frequency, the structure of the flow field around the particle changes significantly, whereas an increasing energy density leads only to increasing the strength of the flow field.   
The propulsion velocity of the particles was found to be approximately proportional to the particle size and acoustic energy density, but to have a nonmonotonic dependence on the frequency with a maximum at about $1\,\mathrm{MHz}$.

As particle size, acoustic energy density, and sound frequency are fundamental parameters of all systems involving ultrasound-propelled particles, our results are highly relevant for the ongoing research on this type of artificial, motile particles. For example, the results can help to plan future experiments and to develop future applications of acoustically propelled nano- and microparticles. 
The observation that the propulsion is maximal for a frequency of about $1\,\mathrm{MHz}$ is particularly beneficial for applications of acoustically propelled particles in medicine since this frequency is sufficiently high to allow precise directing and structuring of the acoustic field and sufficiently low to reach a large penetration depth in tissue \cite{Wells1975}.
Our results can also strongly accelerate theoretical research on such particles, since knowing the strength and direction of the acoustic propulsion for a given system allows to use an effective instead of a direct description of the propulsion and thus to describe the particles' motion on several orders of magnitude larger time scales in analytical modeling and computer simulations \cite{WittkowskiL2012,tenHagenWTKBL2015,BickmannW2020twoD,BickmannW2020b}. 
 
Future research should continue our work by studying the influence of other system parameters on the acoustic propulsion. A particularly important parameter that should be varied in future research is the viscosity of the fluid that surrounds the particles. Up to now, only a few experiments have addressed the acoustic propulsion in fluids with different viscosities \cite{GarciaGradillaEtAl2013,WuEtAl2014,EstebanEtAl2018,GaoLWWXH2019,WangGZLH2020} and they were not able to vary the viscosity independently from other parameters of the system.

\section{\label{methods}Methods}
Our methods are similar to those described in Ref.\ \cite{VossW2020}, which have been proven to be successful. 
The methods mainly consist of direct computational fluid dynamics simulations, which are based on numerically solving the compressible Navier-Stokes equations.
In contrast to other numerical approaches that have been used to study acoustically propelled particles \cite{KaynakONNLCH2016,AhmedBJPDN2016,Zhou2018,SabrinaTABdlCMB2018,WangGWSGXH2018,TangEtAl2019,RenEtAl2019}, our approach solves the full compressible Navier-Stokes equations and does not involve a perturbative expansion, which allows for a higher accuracy. 

\begin{figure}[htb]
\centering
\includegraphics[width=\linewidth]{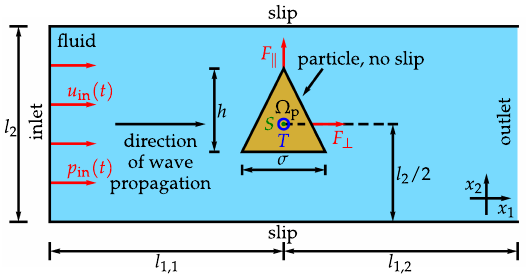}%
\caption{\label{fig:fig7}Setup for the simulations.} 
\end{figure}

The setup for our simulations is shown in Fig.\ \ref{fig:fig7}.
The system includes a fluid-filled rectangular domain with width $l_{1,1}+l_{1,2}$ and height $l_2=\SI{200}{\micro\metre}$, where we choose water as the fluid. 
For convenience, we choose the width to be parallel to the $x_1$-axis and the height to be parallel to the $x_2$-axis of a Cartesian coordinate system.  
At the left edge of the domain, which shall constitute an inlet for the ultrasound wave, we impose time-dependent boundary conditions that correspond to a planar ultrasound wave entering the system.
For this purpose, we prescribe a time-dependent velocity $\ws_{\mathrm{in}}(t)=\Delta u \sin(2\pi f t)$ and pressure $p_{\mathrm{in}}(t)=\Delta p \sin(2\pi f t)$, where $t$ denotes time, $\Delta u=\Delta p/(\rho_0 c_\mathrm{f})$ is the flow velocity amplitude of the entering wave, and $\Delta p$ is its pressure amplitude. 
The water shall initially be at standard temperature $T_0=\SI{293.15}{\kelvin}$ and standard pressure $p_0=\SI{101325}{\pascal}$ so that we set the initial mass density of the water as $\rho_0=\SI{998}{\kilogram\,\metre^{-3}}$, its sound velocity as $c_\mathrm{f}=\SI{1484}{\metre\,\second^{-1}}$, 
its shear viscosity as $\nu_\mathrm{s}=\SI{1.002}{\milli\pascal\,\second}$, and its bulk viscosity as $\nu_\mathrm{b}=\SI{2.87}{\milli\pascal\,\second}$. 
Moreover, we assume that the water is initially at rest, i.e., at $t=0$ it has the vanishing velocity field $\ww_0=\vec{0}\,\SI{}{\metre\,\second^{-1}}$.
In this study, we consider pressure amplitudes of $\Delta p \in \left[0.1,10\right]\Delta p_\mathrm{R}$ with the reference pressure amplitude $\Delta p_\mathrm{R}=\SI{10}{\kilo\pascal}$ and ultrasound frequencies of $f\in \left[\SI{0.5}{},\SI{10}{}\right]f_\mathrm{R}$ with the reference frequency $f_\mathrm{R}=\SI{1}{\MHz}$. 
The reference values are chosen such that they are consistent with previous work \cite{VossW2021}. 
Furthermore, the reference pressure amplitude $\Delta p_\mathrm{R}$ corresponds to an acoustic energy density $E$ (see below for an equation for $E$) that is considered to be harmless and has been approved for diagnostic applications of ultrasound \cite{BarnettEtAl2000}, and the considered frequencies are similar to those used in many experiments that have been reported in the literature \cite{WangCHM2012,GarciaGradillaEtAl2013,AhmedEtAl2013,WuEtAl2014,BalkEtAl2014,GarciaGradillaSSKYWGW2014,SotoWGGGLKACW2016,AhmedBJPDN2016,AhmedWBGHM2016,ZhouZWW2017,SabrinaTABdlCMB2018,TangEtAl2019,DumyJMBGMHA2020}. 
 
At the lower and upper edges of the simulation domain, we prescribe slip boundary conditions. 
The traveling ultrasound wave entering the system at the inlet will then propagate parallel to the $x_1$-axis towards the right edge of the domain, which we choose as outlet so that the ultrasound wave can leave the simulation domain. 
After some time, the wave reaches a cone-shaped particle with a variable diameter $\sigma\in\left[\SI{0.1}{},10\right]\sigma_\mathrm{R}$, where the reference diameter $\sigma_\mathrm{R}=2^{-1/2}\SI{}{\micro\metre}$ is again chosen to be consistent with previous work \cite{VossW2021}. 
The particle's height is set to $h=\sigma$ so that, through $\sigma$, only the size of the particle but not its aspect ratio is varied and the aspect ratio is consistent with the choice in Ref.\ \cite{VossW2021}. 
We position the particle in the simulation domain such that its center of mass $\mathrm{S}$ has a distance $l_{1,1}$ from the inlet and is vertically centered.  
The orientation of the particle, which we describe by the orientation of the particle's axis of symmetry, is chosen so that it is parallel to the $x_2$-axis. 
At the boundary of the particle domain $\Omega_\mathrm{p}$, we prescribe no-slip boundary conditions. 

When the ultrasound, whose acoustic energy density is given by $E=\Delta p^2/(2 \rho_0 c_{\mathrm{f}}^2)\in\left[0.23,2275\right]\SI{}{\milli\joule\,\metre^{-3}}$ and set through the pressure amplitude $\Delta p$, interacts with the particle, a time-averaged, ultrasound-induced propulsion force $\vec{F}$ and torque $T$ are exerted on the center of mass $\mathrm{S}$ of the particle. The propulsion force can be split into components $F_\parallel$ and $F_\perp$ parallel and perpendicular to the particle orientation, respectively. 
After the interaction with the particle, the ultrasound wave propagates further and eventually can reach the outlet, which is at a distance of $l_{1,2}$ from $\mathrm{S}$. 
Analogous to Refs.\ \cite{VossW2020,VossW2021,VossW2022orientation}, we choose the width $l_{1,1}+l_{1,2}$ of the simulation domain so that $l_{1,1}=\lambda(f_\mathrm{R})/4$, where $\lambda(f)=c_\mathrm{f}/f$ is the wavelength of the ultrasound, and so that $l_{1,1}+l_{1,2}$ is a multiple of $\lambda(f)/2$. 
In addition, we demand that $l_{1,2}$ is the smallest value that fulfills also the condition $l_{1,2}\geqslant \SI{100}{\micro\metre}$, to restrict the simulation domain to a reasonable size while ensuring a sufficiently large distance of the particle from the outlet. 
  
To calculate the propulsion force $\vec{F}$ and torque $T$, we first simulate the propagation of the ultrasound wave and its interaction with the particle by numerically solving the continuity equation for the mass-density field of the fluid, the compressible Navier-Stokes equations, and a linear constitutive equation for the fluid's pressure field. 
For this purpose, we use the finite volume software package OpenFOAM \cite{WellerTJF1998}.

From the velocity and pressure fields of the fluid, we then calculate the time-dependent force and torque that act on the particle in the laboratory frame through appropriate integrals of the stress tensor $\Sigma$ over the particle surface. 
The time-dependent force and torque are given by $\vec{F}^{(p)}+\vec{F}^{(v)}$ and $T^{(p)}+T^{(v)}$, respectively, where the pressure component (superscript \ZT{$(p)$}) and viscous component (superscript \ZT{$(v)$}) are given by \cite{LandauL1987}
{\allowdisplaybreaks\begin{align}%
F^{(\alpha)}_{i} &= \sum^{2}_{j=1} \int_{\partial\Omega_{\mathrm{p}}} \!\!\!\!\!\!\! \Sigma^{(\alpha)}_{ij}\,\dif A_{j}, 
\label{eq:F}\\%
T^{(\alpha)} &= \sum^{2}_{j,k,l=1} \int_{\partial\Omega_{\mathrm{p}}} \!\!\!\!\!\!\! \epsilon_{3jk}(x_j-x_{\mathrm{p},j})\Sigma^{(\alpha)}_{kl}\,\dif A_{l} 
\label{eq:T}%
\end{align}}%
with $\alpha\in\{p,v\}$. The symbols $\Sigma^{(p)}$ and $\Sigma^{(v)}$ denote the pressure component and the viscous component of the stress tensor $\Sigma$, respectively, $\dif\vec{A}(\vec{x})=(\dif A_{1}(\vec{x}),\dif A_{2}(\vec{x}))^{\mathrm{T}}$ is the normal, outwards oriented surface element of the particle-domain boundary $\partial\Omega_{\mathrm{p}}$ at position $\vec{x}\in\partial\Omega_{\mathrm{p}}$, $\epsilon_{ijk}$ denotes the Levi-Civita symbol, and $\vec{x}_{\mathrm{p}}$ is the position of $\mathrm{S}$.

Since the time-dependent force and torque converge slowly towards a stationary state, we calculate the time-averaged, stationary force $\vec{F}$ and torque $T$ acting on the particle by locally averaging over one period of the ultrasound wave and extrapolating towards $t \to \infty$ using the procedure described in Ref.\ \cite{VossW2020}. 
We decompose the force as $\vec{F}=\vec{F}_p+\vec{F}_v$ into a pressure component $\vec{F}_p=\langle\vec{F}^{(p)}\rangle$ and viscous component $\vec{F}_v=\langle\vec{F}^{(v)}\rangle$, where $\langle\cdot\rangle$ denotes the average over time.
Similarly, we decompose the torque as $T=T_p + T_v$ with pressure component $T_p=\langle T^{(p)}\rangle$ and viscous component $T_v=\langle T^{(v)}\rangle$. 

We are also interested in the translational velocity $\vec{v}$ and angular velocity $\omega$ of the particle that correspond to $\vec{F}$ and $T$. 
$\vec{v}$ and $\omega$ can be calculated from $\vec{F}$ and $T$ through the Stokes law \cite{HappelB1991}
\begin{equation}
\vec{\mathfrak{v}}=\frac{1}{\nu_\mathrm{s}}\boldsymbol{\mathrm{H}}^{-1}\,\vec{\mathfrak{F}}. 
\label{eq:velocity}%
\end{equation}
For a more compact notation, we use here the translational-angular velocity vector $\vec{\mathfrak{v}}=(\vec{v},\omega)^{\mathrm{T}}$ and the force-torque vector $\vec{\mathfrak{F}}=(\vec{F},T)^{\mathrm{T}}$. 
$\boldsymbol{\mathrm{H}}^{-1}$ is the inverse of the hydrodynamic resistance matrix
\begin{equation}
\boldsymbol{\mathrm{H}}=
\begin{pmatrix}
\boldsymbol{\mathrm{K}} & \boldsymbol{\mathrm{C}}^{\mathrm{T}}_{\mathrm{S}} \\
\boldsymbol{\mathrm{C}}_{\mathrm{S}} & \boldsymbol{\Omega}_{\mathrm{S}} 
\end{pmatrix} 
\label{eq:H}%
\end{equation}
with the submatrices $\boldsymbol{\mathrm{K}}$, $\boldsymbol{\mathrm{C}}_{\mathrm{S}}$, and $\boldsymbol{\Omega}_{\mathrm{S}}$. 
The latter two submatrices depend on a reference point, which we choose here as the center of mass $\mathrm{S}$, as denoted by a subscript $\mathrm{S}$.
 
Since $\boldsymbol{\mathrm{H}}$ corresponds to three spatial dimensions, whereas we perform simulations in two spatial dimensions to keep the computational effort acceptable, we assume a thickness of $\sigma$ of the particle in the third dimension when calculating $\boldsymbol{\mathrm{H}}$.   
For a particle with the reference diameter $\sigma_\mathrm{R}$, this results in 
\begin{align}
\boldsymbol{\mathrm{K}} &= \begin{pmatrix}
\SI{7.74}{\micro\metre} & 0 & 0\\
0 & \SI{7.48}{\micro\metre} & 0 \\
0 & 0 & \SI{7.16}{\micro\metre}
\end{pmatrix},
\label{eq:K}
\\
\boldsymbol{\mathrm{C}}_{\mathrm{S}} &= \begin{pmatrix}
0 & 0 & \SI{0.05}{\micro\metre^2}\\
0 & 0 & 0 \\
\SI{-0.11}{\micro\metre^2} & 0 & 0
\end{pmatrix},
\\
\boldsymbol{\Omega}_{\mathrm{S}} &= \begin{pmatrix}
\SI{1.81}{\micro\metre^3} & 0 & 0\\
0 & \SI{1.69}{\micro\metre^3} & 0 \\
0 & 0 & \SI{1.73}{\micro\metre^3}
\end{pmatrix}.
\label{eq:Omega}%
\end{align}
We then neglect the contributions $\mathrm{K_{33}}$, $\mathrm{C_{13}}$, $\mathrm{\Omega_{11}}$, and $\mathrm{\Omega_{22}}$, which correspond to the lower and upper surfaces of the particle, and can then use the three-dimensional versions of Eqs.\ \eqref{eq:F}-\eqref{eq:velocity}.
The matrix $\boldsymbol{\mathrm{H}}$ can, e.g., be calculated with the software \texttt{HydResMat} \cite{VossW2018,VossJW2019}. 
It is not necessary to calculate $\boldsymbol{\mathrm{H}}$ for each value of $\sigma$. 
Since only the size of the particles is varied in this study, whereas the qualitative shape is unchanged, it is quite easily possible to calculate $\boldsymbol{\mathrm{H}}$ for some value of $\sigma$ from the matrix $\boldsymbol{\mathrm{H}}$ corresponding to $\sigma_\mathrm{R}$ (see Ref.\ \cite{VossW2018} for details):
\begin{align}
\boldsymbol{\mathrm{K}}(\sigma) &= \boldsymbol{\mathrm{K}}(\sigma_\mathrm{R})\frac{\sigma}{\sigma_\mathrm{R}},\label{eq:Ksigma}\\
\boldsymbol{\mathrm{C}}_{\mathrm{S}}(\sigma) &= \boldsymbol{\mathrm{C}}_{\mathrm{S}}(\sigma_\mathrm{R})\bigg(\frac{\sigma}{\sigma_\mathrm{R}}\bigg)^2,\label{eq:Csigma}\\
\boldsymbol{\Omega}_{\mathrm{S}}(\sigma) &= \boldsymbol{\Omega}_{\mathrm{S}}(\sigma_\mathrm{R})\bigg(\frac{\sigma}{\sigma_\mathrm{R}}\bigg)^3.\label{eq:Omegasigma}
\end{align}
From $\boldsymbol{\mathrm{H}}$, one can also calculate the particle's diffusion tensor $\mathcal{D}=(k_\mathrm{B} T_0 / \nu_\mathrm{s}) \boldsymbol{\mathrm{H}}^{-1}$ with the Boltzmann constant $k_\mathrm{B}$. The particle's rotational diffusion coefficient, corresponding to rotation in the $x_1$-$x_2$ plane, is then given by $D_\mathrm{R}=(\mathcal{D})_{66}$. 

In the main part of this article, we discuss the values of the components of $\vec{F}$ and $\vec{v}$ that correspond to the directions parallel and perpendicular to the particle's orientation, respectively. 
The parallel component of $\vec{F}$ is given by $F_\parallel=(\vec{F})_2=F_{\parallel,p}+F_{\parallel,v}$ with pressure component $F_{\parallel,p}=(\langle\vec{F}^{(p)}\rangle)_2$ and viscous component $F_{\parallel,v}=(\langle\vec{F}^{(v)}\rangle)_2$. 
Analogously, the perpendicular component of $\vec{F}$ is given by $F_\perp = (\langle\vec{F}_\perp\rangle)_1=F_{\perp,p}+F_{\perp,v}$ with pressure component $F_{\perp,p}=(\langle\vec{F}^{(p)}\rangle)_1$ and viscous component $F_{\perp,v}=(\langle\vec{F}^{(v)}\rangle)_1$. 
For $\vec{v}$, the parallel component is obtained as $v_{\parallel}=(\vec{v})_2$ and the perpendicular component is given by $v_{\perp}=(\vec{v})_1$.

To assess the type of motion of a particle, one can compare its rotational Brownian motion with its translational and rotational propulsion \cite{VossW2021}. 
First, if the translational and rotational propulsions are weak compared to the Brownian rotation, there is only random motion.
On the other hand, if translational or rotational (which can align the orientation of the particle \cite{VossW2022orientation}) propulsion is strong compared to the Brownian rotation, there is directional motion.
Second, the particle can attain random orientations if the particle's angular propulsion velocity is small compared to the Brownian rotation. 
On the other hand, if the angular propulsion is strong compared to the Brownian rotation, the orientation of the particle is dominated by the propulsion torque so we observe guided motion.
Since the particle's propulsion is proportional to the acoustic energy density $E$ (see Section \ref{results}), the type of motion depends on $E$ as follows \cite{VossW2021}: 
\begin{itemize}
\item ${E < \min\{E_\mathrm{dir},E_\mathrm{gui}\}}$:\quad Random motion 
\item ${E > \min\{E_\mathrm{dir},E_\mathrm{gui}\}}$:\quad Directional motion  
\item ${E < E_\mathrm{gui}}$:\qquad\qquad\qquad Random orientation
\item ${E > E_\mathrm{gui}}$:\qquad\qquad\qquad Guided motion
\end{itemize}
Here, the energy density thresholds are given by \cite{VossW2021}
\begin{align}
E_\mathrm{dir} &= \frac{\sigma D_\mathrm{R} E_\mathrm{R}}{|v_\parallel|}, \label{eq:E_dir_lim} \\
E_\mathrm{gui} &= \frac{\pi D_\mathrm{R} E_\mathrm{R}}{2|\omega|} \label{eq:E_Br_lim}
\end{align}
with the reference acoustic energy density $E_\mathrm{R}=\SI{22.7}{\milli\joule\,\metre^{-3}}$. 

\begin{table*}[htb]
\centering
\caption{\label{tab:Parameters}Summary of the parameters that are relevant for our simulations. Their values are chosen similar as in Ref.\ \cite{VossW2020}. The values of $c_\mathrm{f}$ and $\rho_0$ correspond to quiescent water at normal temperature $T_0$ and normal pressure $p_0$.}%
\begin{ruledtabular}
\begin{tabular}{@{}l@{}c@{}c@{}}%
\textbf{Name} & \textbf{Symbol} & \textbf{Value}\\
\hline
Reference diameter & $\sigma_\mathrm{R}$ & $2^{-1/2}\,\SI{}{\micro\metre}$\\
Particle diameter & $\sigma$ & $\SI{0.1}{}$-$\SI{10}{}\,\sigma_\mathrm{R}$\\
Particle height & $h$ & $\sigma$\\
Reference frequency & $f_\mathrm{R}$ & $\SI{1}{\MHz}$\\
Sound frequency & $f$ & $\SI{0.5}{}$-$\SI{10}{}\,f_\mathrm{R}$\\
Speed of sound & $c_\mathrm{f}$ & $\SI{1484}{\metre\,\second^{-1}}$\\
Time period of sound & $\tau=1/f$ & $\SI{0.1}{}$-$\SI{2}{\micro\second}$\\
Wavelength of sound & $\lambda=c_\mathrm{f}/f$ & $\SI{0.1484}{}$-$\SI{2.968}{\milli\metre}$\\
Temperature of fluid & $T_0$ & $\SI{293.15}{\kelvin}$\\
Mean mass density of fluid & $\rho_0$ & $\SI{998}{\kilogram\,\metre^{-3}}$\\
Mean pressure of fluid & $p_{0}$ & $\SI{101325}{\pascal}$ \\
Initial velocity of fluid & $\ww_{0}$ & $\vec{0}\,\SI{}{\metre\,\second^{-1}}$ \\
Reference sound pressure amplitude & $\Delta p_\mathrm{R}$ & $\SI{10}{\kilo\pascal}$ \\
Sound pressure amplitude & $\Delta p$ & $0.1$-$10\,\Delta p_\mathrm{R}$\\
Flow velocity amplitude & $\Delta u=\Delta p / (\rho_0 c_\mathrm{f})$ & $\SI{0.68}{}$-$\SI{67.52}{\milli\metre\,\second^{-1}}$ \\
Reference acoustic energy density & $E_\mathrm{R}=\Delta p_\mathrm{R}^2/(2 \rho_0 c_{\mathrm{f}}^2)$ & $\SI{22.7}{\milli\joule\,\metre^{-3}}$\\
Acoustic energy density & \hspace*{-0mm}$E=\Delta p^2/(2 \rho_0 c_{\mathrm{f}}^2)$\hspace*{-0mm} & $0.23$-$\SI{2275}{\milli\joule\,m^{-3}}$\\
Shear/dynamic viscosity of fluid & $\nu_{\mathrm{s}}$ & $\SI{1.002}{\milli\pascal\,\second}$ \\
Bulk/volume viscosity of fluid  & $\nu_{\mathrm{b}}$ & $\SI{2.87}{\milli\pascal\,\second}$ \\
Inlet-particle distance & $l_{1,1}$ & $\SI{371}{\micro\metre}$ \\
Particle-outlet distance & $l_{1,2}$ & $\geq\SI{100}{\micro\metre}$\\
Domain width & $l_2$ & $\SI{200}{\micro\metre}$\\
Mesh-cell size & $\Delta x$ & $\SI{7.5}{\nano \metre}$-$\SI{1}{\micro \metre}$ \\
Time-step size & $\Delta t$ & $1$-$\SI{10}{\pico \second}$\\
Simulation duration & $t_{\mathrm{max}}$ & $\geqslant 500\tau$ \\
Euler number & $\mathrm{Eu}$ & 
$\SI{2.20}{\cdot10^4}$-$\SI{2.20}{\cdot10^6}$\\ 
Helmholtz number & $\mathrm{He}$ & $\SI{4.76}{\cdot10^{-5}}$-$\SI{4.76}{\cdot10^{-3}}$ \\
Bulk Reynolds number &  $\mathrm{Re}_\mathrm{b}$ & $\SI{1.66}{\cdot10^{-4}}$-$\SI{1.66}{\cdot10^{-2}}$\\
Shear Reynolds number &  $\mathrm{Re}_\mathrm{s}$ & $\SI{4.76}{\cdot10^{-4}}$-$\SI{4.76}{\cdot10^{-2}}$\\
Particle Reynolds number &  $\mathrm{Re}_\mathrm{p}$ & $<10^{-5}$
\end{tabular}%
\end{ruledtabular}%
\end{table*}

Before solving the system of equations governing our simulations numerically, we nondimensionalize it. 
This leads to four dimensionless numbers, being the Euler number Eu, which corresponds to the pressure amplitude of the ultrasound wave, the Helmholtz number $\mathrm{He}$, which corresponds to the frequency of the wave, a Reynolds number  $\mathrm{Re}_\mathrm{b}$, which corresponds to the bulk viscosity, and a Reynolds number $\mathrm{Re}_\mathrm{s}$, which corresponds to the shear viscosity. 
With the parameter values listed in Tab.\ \ref{tab:Parameters}, these dimensionless numbers obtain the following values:
\begin{align}
\mathrm{Eu}&=\frac{\Delta p}{\rho_0 \Delta u^2}\approx \SI{2.20}{\cdot10^4}\text{-}\SI{2.20}{\cdot10^6},\\
\mathrm{He}&= \frac{f\sigma}{c_\mathrm{f}}\approx \SI{4.76}{\cdot10^{-5}}\text{-}\SI{4.76}{\cdot10^{-3}},\\
\mathrm{Re}_\mathrm{b}&=\frac{\rho_0 \Delta u \sigma}{\nu_\mathrm{b}}\approx \SI{1.66}{}\cdot10^{-4}\text{-}\SI{1.66}{\cdot10^{-2}},\\
\mathrm{Re}_\mathrm{s}&=\frac{\rho_0 \Delta u \sigma}{\nu_\mathrm{s}}\approx \SI{4.76}{\cdot10^{-4}}\text{-}\SI{4.76}{\cdot10^{-2}}.
\end{align}
One can also define the Reynolds number
\begin{align}
\mathrm{Re}_\mathrm{p}=\frac{\rho_0 \sigma}{\nu_\mathrm{s}} 
\sqrt{v_\parallel^2+v_\perp^2}
<10^{-5},
\end{align}
which characterizes the particle motion through the fluid.
See Ref.\ \cite{VossW2021} for a more detailed discussion of the meaning of the dimensionless numbers. 
 
When solving the field equations describing the dynamics of the fluid with the finite volume method, we use a structured, mixed rectangular-triangular mesh with about $300,000$-$800,000$ cells. This mesh has a very small cell size $\Delta x$ close to the particle and a larger $\Delta x$ far away from the particle. 
For the time integration, we use an adaptive time-step method and ensure that the time-step size $\Delta t$ is always sufficiently small such that the Courant-Friedrichs-Lewy number fulfills the condition 
\begin{align}
C = c_\mathrm{f} \frac{\Delta t}{\Delta x} < 1 .
\end{align}
Our simulations run from $t=0$ to $t = t_\mathrm{max} \geqslant 500/f$.  
With the chosen settings, an individual simulation run costs typically $36,000$ CPU core hours of computation time on current hardware. Due to the larger domain width and period for lower frequencies, the computational expense was higher for simulations with lower frequencies. For example, a simulation with $f=\SI{0.5}{\MHz}$ required 144,000 CPU core hours. 

Table \ref{tab:Parameters} summarizes the parameters that are relevant for our simulations. Their values are in line with those chosen in Ref.\ \cite{VossW2020}.

\section*{Data availability}
The raw data corresponding to the figures shown in this article are available as Supplementary Material \cite{SI}.

\section*{Conflicts of interest}
There are no conflicts of interest to declare.

\begin{acknowledgments}
We thank Patrick Kurzeja for helpful discussions. 
R.W.\ is funded by the Deutsche Forschungsgemeinschaft (DFG, German Research Foundation) -- WI 4170/3. 
The simulations for this work were performed on the computer cluster PALMA II of the University of M\"unster. 
\end{acknowledgments}

\nocite{apsrev41Control}
\bibliographystyle{apsrev4-1}
\bibliography{control,refs}
	
\end{document}